\documentclass[fleqn,usenatbib]{mnras}

\usepackage{newtxtext,newtxmath}

\usepackage[T1]{fontenc}

\DeclareRobustCommand{\VAN}[3]{#2}
\let\VANthebibliography\thebibliography
\def\thebibliography{\DeclareRobustCommand{\VAN}[3]{##3}\VANthebibliography}

\usepackage{graphicx}	
\usepackage{amsmath}	
\usepackage{multirow}
\usepackage{float}

\title[Metallicity dependence of disc properties]{The statistical properties of protostellar discs and their dependence on metallicity}

\author[D. Elsender \& M. R. Bate]{
Daniel Elsender,$^{1}$\thanks{E-mail: de296@exeter.ac.uk (DE)}
Matthew R. Bate$^{1}$
\\
$^{1}$ School of Physics and Astronomy, University of Exeter, Stocker Road, Exeter EX4 4QL, UK
}

\date{Accepted October 2021}

\pubyear{2021}

\begin{document}
\label{firstpage}
\pagerange{\pageref{firstpage}--\pageref{lastpage}}
\maketitle

\begin{abstract}
We present the analysis of the properties of large samples of protostellar discs formed in four radiation hydrodynamical simulations of star cluster formation. The four calculations have metallicities of 0.01, 0.1, 1 and 3 times solar metallicity. The calculations treat dust and gas temperatures separately and include a thermochemical model of the diffuse interstellar medium. We find the radii of discs of bound protostellar systems tend to decrease with decreasing metallicity, with the median characteristic radius of discs in the 0.01 and 3 times solar metallicity calculations being $\approx20$ and $\approx65$ au, respectively.  Disc masses and radii of isolated protostars also tend to decrease with decreasing metallicity. We find that the circumstellar discs and orbits of bound protostellar pairs, and the two spins of the two protostars are all less well aligned with each other with lower metallicity than with higher metallicity. These variations with metallicity are due to increased small scale fragmentation due to lower opacities and greater cooling rates with lower metallicity, which increase the stellar multiplicity and increase dynamical interactions. 
We compare the disc masses and radii of protostellar systems from the solar metallicity calculation with recent surveys of discs around Class 0 and I objects in the Orion and Perseus star-forming regions. The masses and radii of the simulated discs have similar distributions to the observed Class 0 and I discs.
\end{abstract}




\section{Introduction}

The formation and evolution of protostellar discs is a key element in understanding of both star and planet formation. They form due to the conservation of angular momentum and dissipation of energy during protostellar collapse. Prior to the advent of interferometers with sub-arcsecond resolution at (sub)-millimetre wavelengths there were relatively few direct images of discs (e.g.  \citealt{beckwith_discovery_of_discs_1984ApJ...287..793B, odell_hst_discs_1993ApJ...410..696O, McCODe1996}; see the review of \citealt{McCStaClo2000}). But with radio telescope arrays such as, the Submillimetre Array (SMA), the Atacama Large Millimetre Array (ALMA), and the Very Large Array (VLA), we can now look deep into star-forming regions to resolve many protoplanetary discs. These interferometers have even allowed us to observe discs of embedded Class 0 and I protostars \citep[e.g.][]{lee_2009_class_0_discs,tobin_2012_l1527irs,yen_2013_class_0_discs,segure-cox_protostar_2016ApJ...817L..14S,aso_2017_l1527irs,yen_protostar_obs_2017ApJ...834..178Y,tychoniec_2018_vandam,tobin_vandam_2020ApJ...890..130T}. 

With the currently growing catalogue of observed discs it is a good time to compare observed discs with those produced in hydrodynamical calculations. The first such comparison of properties of a large sample of discs formed in a hydrodynamical simulation of a star cluster formation was \citet{bate_diversity_of_discs_2018MNRAS.475.5618B} who analysed the discs that were formed in the solar-metallicity calculation first published by \citet{bate_2012_10.1111/j.1365-2966.2011.19955.x}. \citeauthor{bate_2012_10.1111/j.1365-2966.2011.19955.x} studied the large diversity of discs (e.g. disc morphologies, evolutionary processes of discs) and the statistical properties of the discs such as their mass, radii, and disc orientations of bound protostellar pairs. This paper extends this first study to examine the effect metallicity has on the properties of discs. 

The radiation hydrodynamical calculations from which we extract the statistics of disc properties were published by \citet{bate_metallicity_2019MNRAS.484.2341B} who studied the statistical properties of protostars and their dependence on metallicity. This surpassed the work of \citet{bate_metallicity_2014MNRAS.442..285B} by including additional physical processes   \citep{bate_keto_10.1093/mnras/stv451} to better model low density gas. It is not just opacity that changes with metallicity. In previous star formation calculations dust and gas temperatures were usually assumed to be identical \citep[e.g.][]{Bate2009b,Offneretal2009,bate_2012_10.1111/j.1365-2966.2011.19955.x,krumholz_star_formation_2012ApJ...754...71K,myers_star_formation_2013ApJ...766...97M,bate_metallicity_2014MNRAS.442..285B,cunningham_star_formation_2018MNRAS.476..771C}. This is a reasonable approximation when the gas density and/or metallicity are high (e.g. hydrogen number density, $n_{\rm H} > 10^5$~cm$^{-3}$ for solar metallicity;  \citealt{burke_gas_grain_interaction_in_ism_1983ApJ...265..223B,goldsmith_thermal_balance_in_dark_cloudS_2001ApJ...557..736G,glover_molecular_gas_necessary_for_star_formation_2012MNRAS.421....9G}). However, at low densities and/or low metallicities the dust and gas temperatures can become uncoupled \citep{omukai_protostellar_collapse_metallicity_2000ApJ...534..809O,tsuribe_dust_fragmentation_of_low_metal_clouds_2006ApJ...642L..61T,dopcke_dust_cooling_on_low_metal_clouds_2011ApJ...729L...3D,nozawa_low_metal_clouds_2012ApJ...756L..35N,chiaki_dust_grain_growth_in_low_metal_cloud_2013ApJ...765L...3C,dopcke_dust_cooling_low_metal_2013ApJ...766..103D} and gas temperatures are typically higher than dust temperatures \citep[e.g.][]{glover_metal_poor_gas_2012MNRAS.426..377G}. Only changing the opacity may poorly model star formation as fragmentation and gas accretion rates depend on gas temperature. Improvements to thermal modeling at low densities and metallicities by combining radiative transfer with a thermochemical model of the diffuse ISM were developed by \citet{bate_keto_10.1093/mnras/stv451}, see Section \ref{sect:method} for further details. 

In this paper, we report the statistical properties of discs from the four radiation hydrodynamical calculations of star cluster formation by \citep{bate_metallicity_2019MNRAS.484.2341B} that employ the radiative transfer/diffuse ISM method of \citet{bate_keto_10.1093/mnras/stv451}. The calculations have identical initial conditions except for their metallicity. They have 0.01, 0.1, 1 and 3 times solar metallicity. As each of the molecular clouds collapse, discs form around protostars and we examine the statistical properties of samples of these discs to determine the dependence of disc properties on metallicity. In Section \ref{sect:main method} we briefly outline the method and initial conditions that were used to perform the calculations. In Section \ref{sect:results} we first present an overview of the previous statistical study of discs by \citet{bate_diversity_of_discs_2018MNRAS.475.5618B}, and we then report our new results. In Section \ref{sect:discussion} we provide a discussion of observed disc properties and compare them with the properties of the discs that form in the solar metallicity calculation. Finally, in Section \ref{sect:concs}, we give our conclusions.


\section{Method}\label{sect:main method}

The radiation hydrodynamical calculations analysed in this paper were previously reported in \citet{bate_metallicity_2019MNRAS.484.2341B}, but they did not analyse protostellar disc properties. For a complete description of the calculations, see \citet{bate_metallicity_2019MNRAS.484.2341B}. Here we only provide a brief summary of the method and the physical processes that were included.

\subsection{Base SPH method}\label{sect:method}

The calculations were performed using the smoothed particle hydrodyanmics (SPH) code, \texttt{sphNG}, based on the original version by \citet{benz_review_1990nmns.work..269B} and \citet{benz_1990ApJ...348..647B}, but substantially modified by \citet{bate_sphng_1995MNRAS.277..362B}, \citet{price_2007MNRAS.374.1347P}, \citet{whitehouse_2005MNRAS.364.1367W}, and \citet{whitehouse_2006MNRAS.367...32W}, and parallelised using both \texttt{OpenMP} and \texttt{MPI}.

Gravitational forces between particles and a particle's nearest neighbours are calculated using a binary tree. The smoothing lengths of particles are allowed to vary in time and space and are set such that the smoothing length of each particle $h=1.2(m/\rho)^{1/3}$, where $m$ and $\rho$ are the SPH particle's mass and density, respectively \citep[see][]{price_2007MNRAS.374.1347P}. A second-order Runge-Kutta-Fehlberg method \citep{fehlberg1969low} is used to integrate the SPH equations, with individual time-steps for each particle \citep{bate_sphng_1995MNRAS.277..362B}. The artificial viscosity given by \citet{morris_1997JCoPh.136...41M} is used with $\alpha_v$ varying between 0.1 and 1 while $\beta_v = 2\alpha_v$ \citep[see][]{price_2005MNRAS.364..384P}.

\subsection{Radiative transfer and diffuse ISM model}

To treat the thermodynamics the calculations used a method developed by \citet[][]{bate_keto_10.1093/mnras/stv451} to combine radiative transfer and a diffuse ISM model. Here we only give a brief overview of the physics involved in the calculations, the reader is directed to that paper for further details.

An ideal gas equation of state is used for the gas pressure $p=\rho T_{\text{g}} \mathcal{R} / \mu$, where $T_{\text{g}}$ is gas temperature, $\mu$ is the mean molecular weight of the gas (initially set to $\mu=2.38$), and $\mathcal{R}$ is the gas constant. Translational, rotational, and vibrational degrees of freedom of molecular hydrogen are taken into account in the thermal evolution. Additionally, molecular hydrogen dissociation and the ionisation of hydrogen and helium are included, with the mass fractions $X=0.70$ and $Y=0.28$ for hydrogen and helium, respectively. The contributions of metals to the equation of state are neglected.

Gas, dust, and radiation fields have separate temperatures with the thermal evolution combining radiative transfer in the flux-limited diffusion, described by \citet{whitehouse_2005MNRAS.364.1367W} and \citet[][]{whitehouse_2006MNRAS.367...32W}, with a diffuse ISM model similar to that described by \citet{glover_metal_poor_gas_2012MNRAS.426..377G} although with much simplified chemical evolution. The dust temperature assumes local thermodynamic equilibrium with the total radiation field and accounts for thermal energy exchanged between the dust and gas during collisions. The calculations used dust-gas collisional energy transfer rates given by \citet{hollenbach_molecule_formation_1989ApJ...342..306H}.

Various heating and cooling mechanisms for gas are implemented. Heating by direct collisions with cosmic rays, indirect heating via hot electrons from dust grains due to the photoeletric effect by photons from the interstellar radiation field (ISRF), and heating from molecular hydrogen forming on dust grains are included. The ISRF used is in the form of \citet{zucconi_dust_temp_2001A&A...376..650Z}, adapted to include the UV component from \citet{draine_photoelectric_1978ApJS...36..595D}. Cooling by electron recombination, atomic oxygen and carbon fine structure cooling, and molecular line cooling are included.

The chemical model of \citet{keto_chemistry_2008ApJ...683..238K} is used to compute the abundances of C+, C, CO, and the depletion of CO on to dust grains. The abundance of atomic oxygen is assumed to scale proportional to $(1-{\rm CO/C})$. The molecular hydrogen formation and dissociation rates used to compute the abundance of atomic and molecular hydrogen are the same as used by \citet{glover_modelling_co_formation_2010MNRAS.404....2G}.

Opacity is set according to the tables of \citet{pollack_opacity_tables_1985Icar...64..471P} at low temperatures when dust is present and it is assumed opacity scales linearly with metallicity. At higher temperatures the tables of \citet{ferguson_opacity_tables_2005ApJ...623..585F} are used with $X=0.70$, covering heavy element abundances from $Z=0$ to $Z=0.1$; the solar abundance is taken to be $\text{Z}_\odot=0.02$. Dust properties may change with different metallicity \citep[e.g.,][]{remy_dust_properties_2014A&A...563A..31R} but these are not taken into account (see \citealt{bate_metallicity_2019MNRAS.484.2341B} for further discussion).

\subsection{Sink particles}\label{sec: sink particles}

During each calculation the protostellar collapse is followed into the second collapse phase caused by the dissociation of molecular hydrogen \citep[][]{larson_1969MNRAS.145..271L}. The timesteps become increasingly small, so sink particles \citep[see][]{bate_sphng_1995MNRAS.277..362B} are inserted after the gas density exceeds $10^{-5}~\text{g cm}^{-3}$. All SPH particles within $r_{\text{acc}} = 0. 5$ au of the densest particle are replaced by a sink particle with the same combined mass and momentum. If an SPH particle comes within $r_{\text{acc}}$, is bound, and has a specific angular momentum less than that required to enter a circular orbit with radius $r_{\text{acc}}$, it is accreted by the sink particle. Because of this circumstellar discs can only be resolved if they have a radius greater than $\approx 1$ au. The angular momentum of the particles accreted is used to calculate the spin of the sink particles but does not have an effect on the rest of the calculation. The sink particles do not contribute to radiative feedback \citep[][provide a detailed discussion on the limitations of sink particle approximation]{bate_2012_10.1111/j.1365-2966.2011.19955.x}. Sink particles merge if they pass within 0.03 au of each other.

\subsection{Initial conditions}\label{sec:int_conds}

For a more complete description of the initial conditions of the four calculations the reader is again directed to \citet[][]{bate_metallicity_2019MNRAS.484.2341B}. The initial density and velocity structure was identical for all four calculations. A uniform density, spherical cloud of gas contained $500~\text{M}_\odot$ with radius 0.404pc with an initial density of $1.2\times10^{-19}~\text{g cm}^{-3}$ (hydrogen number density $n_{\rm H} = 6\times 10^4$~cm$^{-3}$). The initial free fall time of the gas was $t_{\text{ff}}=1.90\times10^{5}~\text{yr}$. An initial supersonic turbulent velocity field was applied to each cloud, in the manner of \citet[][]{Ostriker_turbulence_2001ApJ...546..980O} and \citet[][]{bate_cluster_formation_2003MNRAS.339..577B}. The velocity field is a divergence-free random Gaussian with a power spectrum $P(k)\propto k^{-4}$, where $k$ is the wavenumber, on a $128^3$ uniform grid with particle velocities interpolated from the grid. 

For each calculation the dust is initially in thermal equilibrium with the local ISRF, and the gas is in thermal equilibrium with heating from the ISRF and cosmic rays with cooling provide by atomic and molecular line emission and collisional coupling with the dust. This causes a range of initial temperatures for the gas and dust, with dust being warmest on the outside of the cloud and coolest in the centre. For $Z=3~\text{Z}_\odot$, $T_{\text{dust}}=6.3 - 17$~K, for $Z=\text{Z}_\odot$, $T_{\text{dust}}=7.1 - 17$~K, for $Z=0.1~\text{Z}_\odot$, $T_{\text{dust}}=12 - 17$~K, for $Z=0.01~\text{Z}_\odot$, $T_{\text{dust}}=16-18$~K. The initial gas temperatures vary less with $T_{\text{g}}=9.1-9.8$~K.

Each cloud was modelled by $3.5\times 10^7$ SPH particles, providing enough resolution to resolve the local Jeans mass throughout the calculation, necessary to correctly model fragmentation down to the opacity limit \citep{bate_resolution_1997MNRAS.288.1060B,truelove_resolution_1997ApJ...489L.179T,whitworth_jeans_instab_1998MNRAS.296..442W,boss_jeans_2000ApJ...528..325B,hubber_resolution_2006A&A...450..881H}.

\subsection{Disc characterisation}

The protostellar discs in each calculation continuously evolve due to a variety of processes. Following \citet{bate_diversity_of_discs_2018MNRAS.475.5618B}, we sample the properties of discs at multiple times during the calculation to examine the disc properties statistically. These samples are take every $0.0025 t_{\text{ff}}$ ($\approx480\text{yr}$) for each protostar. The number of instances of discs resulting from each of the four calculations of \citet[][]{bate_metallicity_2019MNRAS.484.2341B} and from the calculation of \citet[][]{bate_2012_10.1111/j.1365-2966.2011.19955.x} are found in the second column in Table \ref{tab:number of discs}. From here on in when referring to discs formed in the \citet{bate_2012_10.1111/j.1365-2966.2011.19955.x} calculation we will usually refer only to \citet{bate_diversity_of_discs_2018MNRAS.475.5618B}.

\subsubsection{Circumstellar discs}

For each protostar (modelled by a sink particle), the SPH gas particles (and other sink particles) are sorted by distance from the sink particle. An SPH particle is considered to be part of the protostellar disc if it has not already been assigned to a disc of a different protostar and if the instantaneous ballistic orbit of the particle has an apastron distance of less than 2000 AU and an eccentricity $e<0.3$. We do this starting with the SPH particle closest to the sink particle. The sensitivity to the choice of eccentricity limit is discussed in \citet[][section 2.3.3]{bate_diversity_of_discs_2018MNRAS.475.5618B}. If these criteria are met, the mass of the particle is added to the mass of the system then the position and velocity of the centre of mass of the system are calculated. This process is repeated with the updated quantities for the next SPH particle. No particles further than 2000 au from the centre of mass are considered -- this distance was chosen as it is larger than the apparent radius of any disc studied by \citet{bate_diversity_of_discs_2018MNRAS.475.5618B}, and also applies to the discs of \citet[][]{bate_metallicity_2019MNRAS.484.2341B}.

If, when moving through the list of particles sorted by distance, we encounter a sink particle (e.g. either as part of a system, or a passing protostar) we record the identity of the sink and the process of adding mass to the disc is stopped. No particles further away than the nearest sink particle are included. We refer to protostars that do not have a companion within 2000 au as being isolated and protostars that have never had a companion within 2000 au as having no encounters. A protostar can be single but not isolated -- if a protostar has another protostar within 2000 au but the two are not bound then it is single protostar.

The above algorithm gives sensible extraction of discs for both the calculation analysed in \citet{bate_diversity_of_discs_2018MNRAS.475.5618B} and those analysed here. It is, however, difficult to separate the disc and the envelope of the protostar and sometimes the algorithm finds low mass `discs' with very large radii. We judge these to not really be discs, rather parts of the infalling envelope. To avoid counting these as discs we exclude any `disc' with a mass $<0.03~\text{M}_\odot$ ($<2100$ SPH particles) and the radius that contains $63\%$ (see below for how this is determined) of this mass is $>300$ au. Additionally, any `disc' that has a radius containing $63\%$ of its mass that is three times larger than the radius containing $50\%$ of its mass is excluded. These cuts reduce the number of instances of circumstellar discs that we use for the statistical analysis of the four calculations of \citet[][]{bate_metallicity_2019MNRAS.484.2341B} (see the third column in Table \ref{tab:number of discs} for the number of instances we use and the number of instances used in the disc analysis by \citealt{bate_diversity_of_discs_2018MNRAS.475.5618B}).

The truncated power-law disc radial surface density profile
\begin{equation}\label{surf_density}
    \Sigma(r)=\Sigma_\text{c}\left(\frac{r}{r_\text{c}}\right)^{-\gamma}\exp{\left[-\left(\frac{r}{r_\text{c}}\right)^{2-\gamma}\right]},
\end{equation}
is often used by observers \citep[e.g.][]{2017A&A...606A..88T, fedele_rings_and_gaps_in_ppd_2017A&A...600A..72F} to fit observed discs.  In this function, $r_\text{c}$ is the characteristic radius of the disc, $\gamma$ is the power-law slope, and $\Sigma_\text{c}/e$ is the gas surface density at $r_\text{c}$. As pointed out by \citet{bate_diversity_of_discs_2018MNRAS.475.5618B} for $\gamma<2$, $r_\text{c}$ is always equal to the radius that contains $(1-1/e)$ of the total disc mass (i.e. $63.2\%$). We note that equation (\ref{surf_density}) only gives sensible profiles for $\gamma<
2$. So if the disc well described by equation (\ref{surf_density}) we obtain $r_\text{c}$ simply by measuring the radius that contains $63.2\%$ of the total disc mass. When referring to a disc radius in this paper we mean $r_\text{c}$ as defined above.
\begin{table*} 
    \centering
    \begin{tabular}{p{2.5cm}|p{2cm}|p{2cm}|p{3cm}}
        \hline
        Calculation & Instances of discs & Instances of discs used in analysis & Instances of isolated discs used in analysis\\ 
        \hline
        Bate 2012 & 11831 & 11281 & 2186\\
        Metallicity 3~Z$_\odot$ & 18172 & 17003 & 3648 \\
        Solar Metallicity & 16048 & 15323 & 2972 \\
        Metallicity 0.1~Z$_\odot$ & 8380 & 8034 & 1779 \\
        Metallicity 0.001~Z$_\odot$ & 4809 & 4585 & 826 \\
        \hline
    \end{tabular}
    \caption{The instances of discs found in each of the four calculations analysed here and the calculation of \citet{bate_2012_10.1111/j.1365-2966.2011.19955.x} that was analysed by \citet{bate_diversity_of_discs_2018MNRAS.475.5618B}. We apply criteria to determine which instances we consider as `real' disc that decreases the number of instances of discs we use in our analysis.  The largest decrease in instances is $6.4\%$ for the $3~\text{Z}_\odot$ calculation. All calculations were run to $1.20~t_{\text{ff}}$ ($\approx 230,000$~yr). We identify fewer discs with decreasing metallicity.}
    \label{tab:number of discs}
\end{table*}

\subsubsection{Circum-multiple discs}
\label{sect: multiple discs}

Many of the protostars in these calculations are found to be in bound multiple systems and thus discs formed in these systems are far more complex than those around a single star. Due to the complexity of these higher order systems we limit our analysis to single, binary, triple, and quadruple systems and to general properties (e.g. disc mass, disc radius, disc/star mass ratios). Systems with an order higher than four that are made up of individual bound systems of order four or less are treat as separate systems. For example, a septuple system consisting of a quadruple system bound to a triple system will be treat as two individual systems. For protostars bound in pairs, either as a binary or as components of hierarchical higher order systems, we also examine the alignments of the circumstellar discs, protostellar spins, and the orbital plane of the pair.

To compute the total disc mass of a system we sum the total mass of all the discs extracted for a system. Determining the characteristic disc radius for a multiple system is not straight forward. For circumstellar discs, we record the radii containing 2, 5, 10, 20, 30, 40, 50, 63.2, 70, 80, 90, 95, 100 percent of the disc mass. To calculate the characteristic radius for a multiple system we loop over all of the component discs, starting with the smallest of the above radii for each disc, and keep a cumulative sum of the mass contained within a given radius. For example, consider a binary system for which the radii containing $2\%$ and $5\%$ for the disc masses are 4 au and 8 au for the primary's disc, and 3 au and 7 au for the secondary's disc, and 50 au and 75 au for the circumbinary disc. We first sum $2\%$ of the secondary's disc mass with $2\%$ of the primary's disc mass, then with $3\%$ ($5\%-2\%$)  of the secondary's disc mass, then with $3\%$ of the primary's disc mass, and so on. The characteristic radius for the system is then the radius at which the cumulative sum first exceeds $63.2\%$ for the total disc mass of the system (again, see \citealt{bate_diversity_of_discs_2018MNRAS.475.5618B} for further details).

The analysis of the protostellar systems uses the same constraints as for circumstellar discs. Any circum-multiple disc which has a total mass <0.03~$\text{M}_\odot$ (<2100 SPH particles) and a radius containing $63.2\%$ of the disc mass that is >300 au is excluded. In addition to this any disc with a characteristic radius more than three times great than the radius containing $50\%$ of the disc mass is also excluded.


\section{The statistical properties of the discs}\label{sect:results}

Here we discuss the statistical properties of the discs. We begin by providing an overview of the results of the disc analysis reported by \citet{bate_diversity_of_discs_2018MNRAS.475.5618B}. Next we consider the properties of circumstellar discs (orbiting just one protostar), discs of isolated protostars, and discs of protostars that have had no encounters (see the definitions below). We then discuss properties of discs in bound protostellar systems, focusing on the mass and radius of these discs. Finally we investigate the orientation angles between discs, orbital planes, and sink particle spins (angular momenta of protostar and inner disc).

\subsection{\citet{bate_diversity_of_discs_2018MNRAS.475.5618B} disc analysis}\label{sect:bate}

In this paper we perform a similar analysis to \citet{bate_diversity_of_discs_2018MNRAS.475.5618B} to gather statistical properties of discs and investigate their dependence on metallicity. We begin by summarising the main finding of \citet{bate_diversity_of_discs_2018MNRAS.475.5618B} to put our results in context, and we discuss the different physics used in the calculations \citep[][]{bate_2012_10.1111/j.1365-2966.2011.19955.x,bate_metallicity_2019MNRAS.484.2341B}. 

The calculation analysed by \citet{bate_diversity_of_discs_2018MNRAS.475.5618B} assumed solar metallicity and employed two-temperature (gas and radiation) radiative transfer in the flux-limited diffusion approximation as developed by \citet{whitehouse_2005MNRAS.364.1367W} and \citet{whitehouse_2006MNRAS.367...32W}. The gas and dust temperatures were assumed to be the same throughout the calculation.  By contrast, the \citet{bate_metallicity_2019MNRAS.484.2341B} calculations used the \citet{bate_keto_10.1093/mnras/stv451} method to combine radiative transfer with a diffuse ISM model.  This method has separate gas and dust temperatures and includes energy transfer between the gas and dust via collisions. The dust temperature is set by assuming it is in local thermodynamic equilibrium with the total radiation field. Initially the density and velocity structure are the same for all of the calculations (both  \citealt{bate_2012_10.1111/j.1365-2966.2011.19955.x} and \citealt{bate_metallicity_2019MNRAS.484.2341B}), but the initial temperatures are different.  The calculation of \citet{bate_2012_10.1111/j.1365-2966.2011.19955.x} had a uniform initial temperature of 10.3~K, while the initial gas and dust temperatures vary both spatially (due to extinction of the ISRF) and due the differing metallicity.  During the evolution of the calculations, it is the low-density and/or low-metallicity gas whose temperatures differ most from the temperatures obtained by \citet{bate_2012_10.1111/j.1365-2966.2011.19955.x}.  For the solar-metallicity calculation of \citet{bate_metallicity_2019MNRAS.484.2341B}, because the initial cloud density is quite high ($n_{\rm H} = 6\times 10^4$~cm$^{-3}$) and the gas and dust are reasonably well coupled thermally by collisions, the gas temperatures are similar to those obtained by \citet{bate_2012_10.1111/j.1365-2966.2011.19955.x} who also assumed solar metallicity, except in the outermost parts of the cloud (where star formation does not occur).  Therefore, it is expected that the discs of the solar metallicity calculation analysed here should have similar properties to the calculation analysed by \citet{bate_diversity_of_discs_2018MNRAS.475.5618B}.

\citet{bate_diversity_of_discs_2018MNRAS.475.5618B} show many images of discs formed in the calculation to demonstrate the diversity of discs.  The calculations of \citealt{bate_metallicity_2019MNRAS.484.2341B} also have a wide diversity of discs (see the mosaic animations published as supplementary information with the paper), but in this paper we only consider their bulk statistical properties. \citet{bate_diversity_of_discs_2018MNRAS.475.5618B}  define an isolated disc as one without a protostellar companion closer than 2000 au. This definition allows for the inclusion of discs that may have been part of multiple systems or had close encounters with other discs, or may become part of a multiple system in the future. Having isolated discs defined this way is consistent with what an observer would see; they would not know the history or future of the disc, only that is is currently isolated. A protostar that has never had an encounter within 2000 au is classified as having no encounters. We categorise discs in the same way to make meaningful comparisons between the calculations. 

\subsubsection{Disc masses and radii}

The analysis done by \citet{bate_diversity_of_discs_2018MNRAS.475.5618B} was the first attempt at protostellar disc population synthesis using a hydrodynamical calculation. Their aims were to show the large diversity of disc types that can be expected around young stars, and to provide statistics on their evolution and properties (e.g. mass, radii, disc alignment). We do not provide a detailed investigation into the evolution of discs -- the same evolutionary processes are involved in sculpting the disc populations and in most cases similar evolutionary trends are found.

\citeauthor{bate_diversity_of_discs_2018MNRAS.475.5618B} initially consider the discs around individual protostars (i.e., circumstellar discs), including isolated discs and discs with no encounters as defined above. They find that most ($\approx70\%$) circumstellar discs are not resolved (i.e. $M_\text{d}<0.01\text{M}_\odot$), largely due to interactions with other protostars or ram-pressure stripping.  Protostars that have had no encounters mostly have resolved discs with the masses of these discs having a clear dependence on the mass of the protostar they are orbiting. The disc mass distributions are subdivided into three protostellar mass ranges: $M<0.1~\text{M}_\odot$, $0.1~\text{M}_\odot\le M<0.3~\text{M}_\odot$, and $M>0.3~\text{M}_\odot$ and the typical disc mass is found to scale approximately linearly with protostellar mass.

Next \citet{bate_diversity_of_discs_2018MNRAS.475.5618B} investigate the statistical properties of the discs of stellar systems. When considering the dependency of total system disc mass on stellar system mass they use the same three mass ranges as above.  Discs in systems that have a higher total mass tend to be more massive until the mass of the system exceeds $\approx 0.5~\text{M}_\odot$, above which the typical total disc mass is found to be more or less independent of the total stellar mass. More than half of very low mass (VLM) ($M<0.1~\text{M}_\odot$) systems have unresolved discs. The discs in the VLM systems tend to be a factor of two times smaller in radius than systems with $0.1~\text{M}_\odot\le M <0.3~\text{M}_\odot$ and a factor of three times smaller than systems with $M>0.3~\text{M}_\odot$. The largest discs tend to be found in multiple systems.

\subsubsection{Disc orientations}

\citet{bate_diversity_of_discs_2018MNRAS.475.5618B} provide a discussion of the relative orientations of discs, orbits, and sink particle spins (this can be thought of as the angular momentum of the star and the inner region of the disc) in bound protostellar pairs. Pairs can be either binary systems or a mutual closest neighbour in a multiple system. To be considered, discs must have a mass of $M_{\text{d}}\ge4.3\times10^{-4}~\text{M}_\odot$. This is equivalent to 30 SPH particles which is enough to calculate the angular momentum vector of the disc.

Discs tend to be more aligned with each other in closer systems, with discs having semi-major axes $\lesssim 100$ au typically being strongly aligned. The discs become more aligned with increasing age, and discs that are part of higher order systems ($>2$ protostars) tend to be more well aligned than discs in binary systems. They suggest this is likely because pairs in higher order systems originate from disc fragmentation more often that discs in binary systems, in which discs originate from either disc fragmentaion or star-disc interactions. The alignment between discs and the orbit of the pair has a weaker alignment for close systems and a stronger alignment in wider systems when compared to the alignment between discs. There is less of a dependence of alignment on age than there was for disc-disc alignment. 

The alignment between sink particle spins of pairs (angular momentum of the protostar and inner disc on the scale of $\lesssim 0.5$ au) is similar to the alignment between circumstellar discs. Spin-spin alignment have the same dependencies on age, separation and multiplicity as for disc-disc alignment. The circumstellar discs and sink particle spins of bound pairs show a tendency for strong alignment. Roughly $50\%$ of protostars have misalignments of more than $30\degr$ between their resolved outer discs and their sink particle spins (i.e., the combination of their protostellar and inner disc angular momenta). The explanation for this is that the outer part of the discs are continually having their orientations changed more quickly than the spins are. There is a lag for the reorientation of spins through accretion from larger scales of the disc. The alignment of sink particle spins and circumstellar discs do not have the same dependencies on age, separation, and multiplicity as the spin-spin alignment (circumstellar discs and protostellar spins are generally much better aligned with each other than the two spins of a pair of protostars).

\subsection{Circumstellar discs}

\begin{figure*}
    \centering
    \includegraphics[scale=0.59]{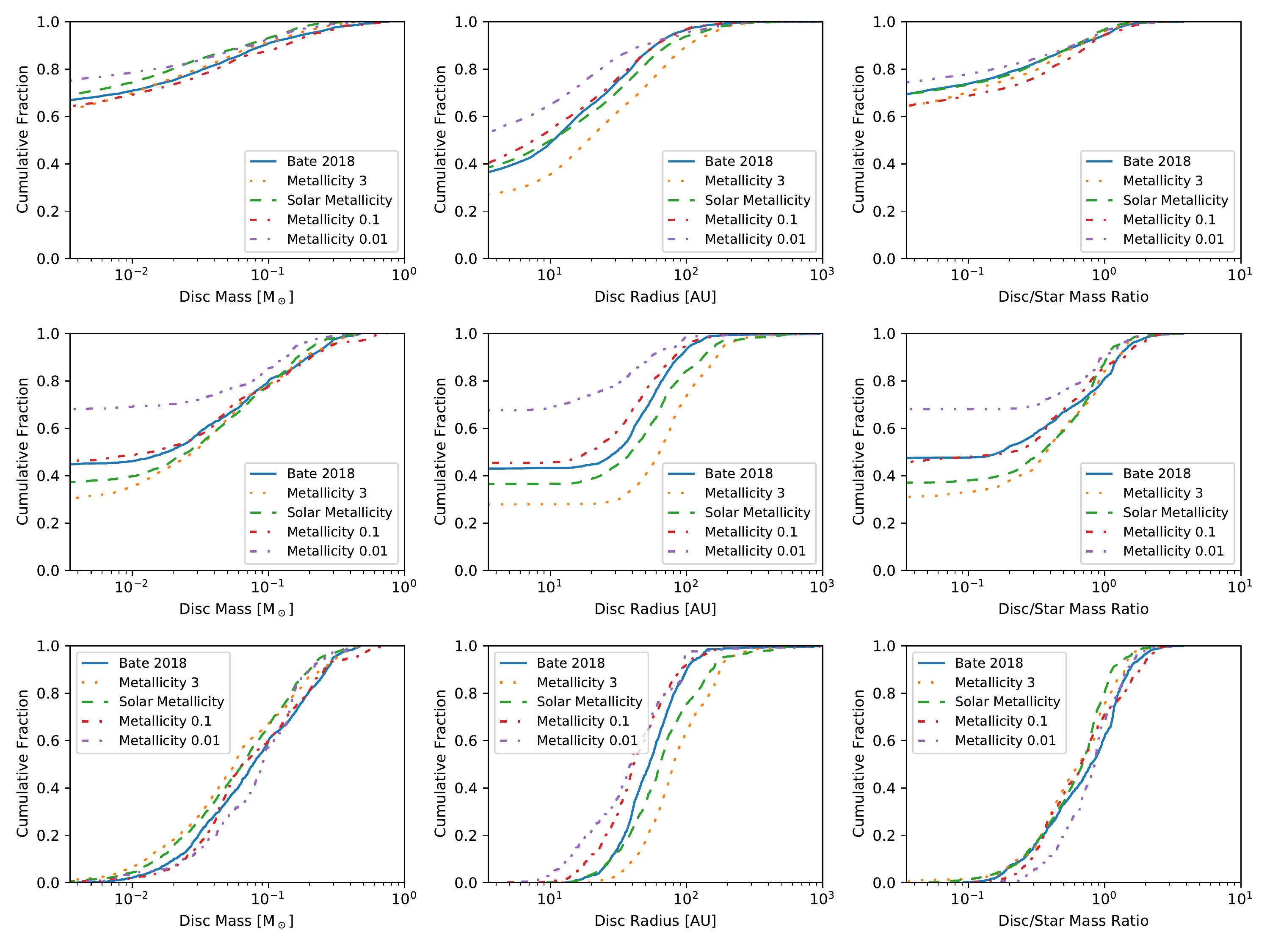}  \vspace{-0.5cm}
    \caption{Cumulative distributions of disc mass (left), characteristic radius (centre), and disc/star mass ratio (right) for circumstellar discs (top), isolated discs (middle), and discs that have never had a companion protostar pass within 2000 au of the protostar that they orbit (bottom). Here we show the comparison between discs found in the four calculations with differing metallicity and discs found in \citet[][]{bate_diversity_of_discs_2018MNRAS.475.5618B}.}
    \label{fig:cdists_disc}
\end{figure*}

In Fig. \ref{fig:cdists_disc} we compare the cumulative distributions of circumstellar disc masses, radii, and disc/star mass ratios from each of the four calculations with different metallicity with the analysis of \citet[][]{bate_diversity_of_discs_2018MNRAS.475.5618B}. The top row gives cumulative distributions for all discs orbiting only one protostar, the middle row gives the corresponding distributions for isolated protostars, and the bottom row gives the distributions for protostars that have never been within 2000 au of another protostar.

We see in the top left panel that most circumstellar discs are unresolved (i.e. $M_\text{d}<0.01~\text{M}_\odot$), however those discs that are resolved generally follow the same trends and there is no consistent trend with metallicity.

In the middle left panel we see that isolated protostars tend to have mostly resolved discs, with the exception of the lowest metallicity calculation, and there is a consistent trend that a higher proportion of protostars have resolved discs with increasing metallicity.  There are also very few massive discs with the lowest metallicity.  For protostars never having had another protostar within 2000 au, the vast majority have resolved discs, since they have not been disrupted or truncated by dynamical interactions with other protostars. The mass distributions of these discs have a weak metallicity dependence such that the low-mass discs ($0.01~{\rm M}_\odot < M_{\rm d} < 0.1~{\rm M}_\odot$) in the lowest metallicity calculation are typically $\approx 50$\% more massive than those in the highest metallicity calculation.  As mentioned above, however, there is also a relative deficit of massive discs in the lowest metallicity calculation ($Z=0.01~{\rm Z}_\odot$).

From the centre column it is very clear that larger discs tend to form with higher metallicity; this is most clear from the bottom centre panel as these no encounter discs are mostly resolved, but the trend is apparent in all of the samples.

As for the cumulative distributions of the ratio between the disc and protostar masses (panels in the right column), there does not appear to be a consistent trend with the metallicity of the molecular cloud, except in the case of the isolated disc sample.  For isolated discs, the cumulative distributions becomes steeper with increasing metallicity (i.e. with low metallicity there are many unresolved discs or discs less massive than the protostar). 

\begin{figure*}
    \centering
    \includegraphics[scale=0.59]{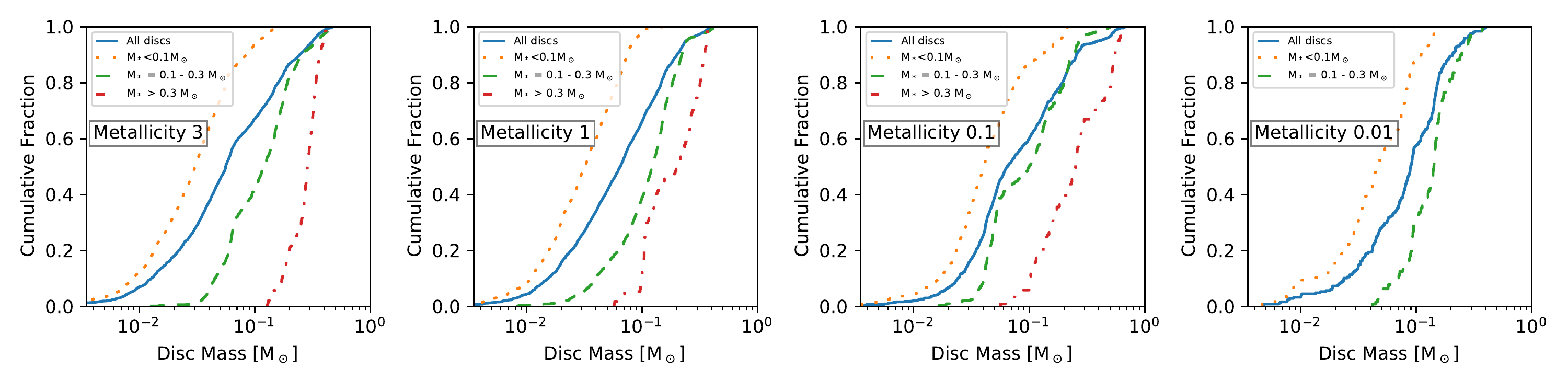}  \vspace{-0.5cm}
    \caption{The cumulative distributions of circumstellar disc mass for protostars that have never had an encounter within 2000 au. The solid line gives the distribution for no encounter discs, and we also provide the distributions for sub-samples in which the protostellar mass lies in the ranges $M<0.1\text{M}_\odot$, $0.1\text{M}_\odot<M<0.3\text{M}_\odot$, and $M>0.3\text{M}_\odot$. Discs orbiting more massive stars tend to be more massive. The masses of no encounter discs tend to increase slightly as metallicity decreases. Note that we do not include the cumulative distribution of disc mass for stars of mass $M>0.3\text{M}_\odot$ from the $Z=0.01{\rm Z}_\odot$ metallicity calculation as there are only two such instances of isolated discs (see Table \ref{tab:number of no encs}).}
    \label{fig:no enc mass}
\end{figure*}

In Fig. \ref{fig:no enc mass} we show the distributions of masses of discs about protostars that have not had an encounter with another protostar within 2000 au. It is clear that disc mass scales with star mass for each metallicity with most disc masses being in the range $\approx$ 0.02-0.2 $\text{M}_\odot$. There is a slightly greater spread of disc mass across the mass bins in the super-solar metallicity calculation than in the lower metallicities. In Table \ref{tab:number of no encs} we give the number of instances of discs about protostars that have had no encounters within 2000 au of an other protostar. We note that there are significantly fewer protostars having no encounters of mass $M>0.3~\text{M}_\odot$ for the lowest metallicity. As discussed in \citet{bate_metallicity_2019MNRAS.484.2341B} the lowest metallicity calculation has the highest multiplicity out of the four calculations, making it less likely for protostars to have no encounters within 2000 au.  This is also why the $Z=0.01~{\rm Z}_\odot$ calculation has the highest fraction of unresolved discs (Fig. \ref{fig:cdists_disc}) -- a greater fraction of stars have close companions that disrupt or truncate their circumstellar discs.  The reason for the higher multiplicity at low metallicity and the above trends of circumstellar disc properties with metallicity is discussed in Section \ref{sect:discussion:opacity}.

\begin{table*}
    \centering
    \begin{tabular}{|c|c|c|c|c|c|}
        \hline
        Calculation & & $M<0.1~\text{M}_\odot$ & $0.1~\text{M}_\odot<M<0.3~\text{M}_\odot$ & $M>0.3~\text{M}_\odot$ & Total \\
        \hline
        \multirow{2}{7em}{Metallicity 3}  & Instances & 1295 & 859 & 195 & 2349 \\
         & Percentage & 55.1\% & 36.6\% & 8.3\% & - \\
         \multirow{2}{7em}{Metallicity 1} & Instances & 896 & 812 & 115 & 1823\\
         & Percentage & 49.2\% & 44.5\% & 6.3\% & - \\
         \multirow{2}{7em}{Metallicity 0.1} & Instances & 359 & 329 & 122 & 810 \\
         & Percentage & 44.3\% & 40.6\% & 15.1\% & - \\
         \multirow{2}{7em}{Metallicity 0.01} & Instances & 118 & 129 & 2 & 249 \\
         & Percentage & 47.4\% & 51.8\% & 0.8\% & - \\
         \hline
    \end{tabular}
    \caption{We provide the numbers of instances of discs orbiting protostars that have never had another protostar within 2000 au for each of the four calculations.  We provide the total number, and the numbers in three bins based on protostellar mass. Alongside these we provide the number of instances in each mass bin as a percentage of the total number of instances of these discs for each calculation. We generally see the proportion of discs decrease as the mass bin increases. This is not case for the lowest metallicity calculation where we see a slight uptick in instances of discs in the intermediate mass bin and almost no instances of discs in the high mass bin. As the metallicity is decreased the number of discs about protostars with no encounters within 2000 au consistently decreases. }
    \label{tab:number of no encs}
\end{table*}

\subsection{Discs of bound systems}

In this section, we discuss the properties of discs found in bound protostellar systems.  These include circumstellar, circumbinary and circum-multiple discs up to and including discs surrounding quadruple systems. 

\subsubsection{Disc masses and radii}

\begin{figure*}
    \centering
    \includegraphics[scale=0.59]{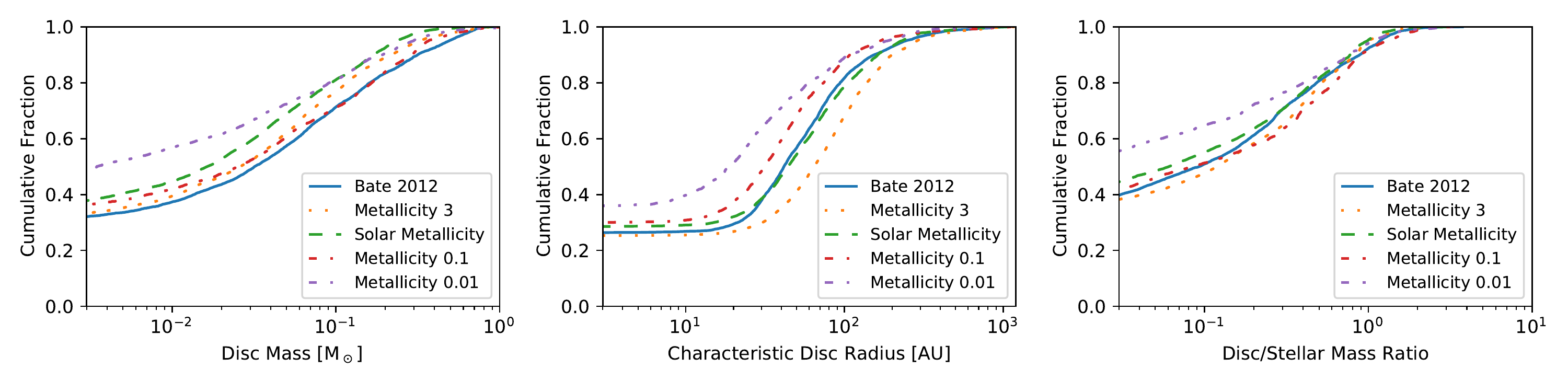}  \vspace{-0.5cm}
    \caption{Cumulative distributions of protostellar system total disc mass (left), characteristic radius (centre), and total disc mass/total stellar mass ratios (right). We compare the cumulative distributions from each of the four metallicity calculations with those from the \citep[][]{bate_2012_10.1111/j.1365-2966.2011.19955.x} calculation.  Note that the solar metallicity calculation has similar distributions for disc radius and disc/star mass ratio to the \citep[][]{bate_2012_10.1111/j.1365-2966.2011.19955.x} calculation (which also assumed solar metallicity), though the system disc masses themselves are typically about a factor of two lower in the newer calculation. }
    \label{fig:cdists_sys}
\end{figure*}

\begin{figure*}
    \centering
    \includegraphics[scale=0.59]{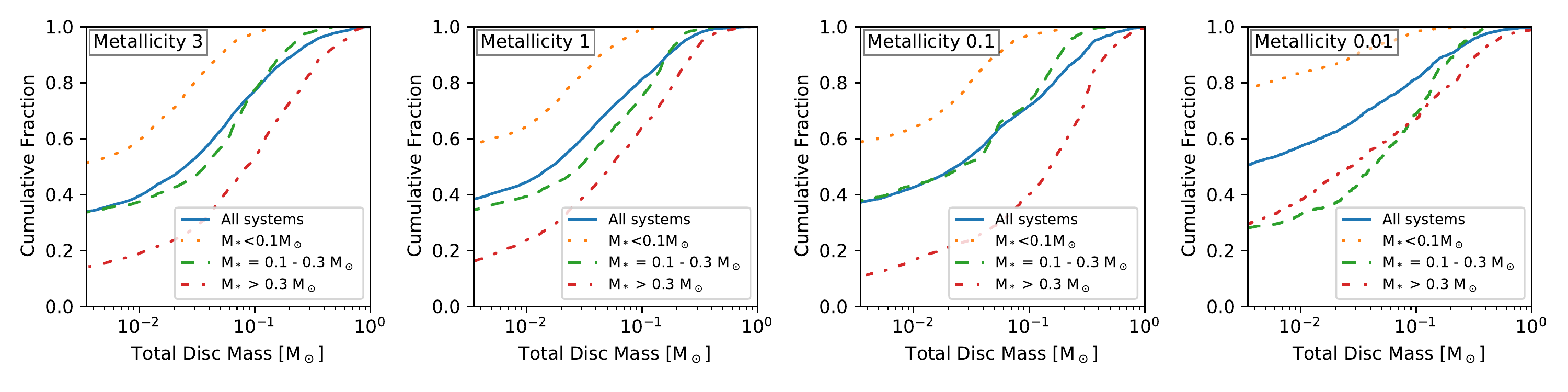}  \vspace{-0.5cm}
    \caption{The cumulative distributions of total disc masses of protostellar systems for each of the four calculations. We give the distributions of all discs, discs about systems of total stellar mass $M<0.1~\text{M}_\odot$, $0.1~\text{M}_\odot<M<0.3~\text{M}_\odot$, and $M>0.3~\text{M}_\odot$. The total disc mass tends to increase as the stellar mass of the system increases, except in the $Z=0.01\text{Z}_\odot$ case where the disc masses in the two higher protostellar mass ranges are similarly distributed.}
    \label{fig:sys_mass_mass}
\end{figure*}

In Fig. \ref{fig:cdists_sys} we compare the same three properties across the calculations as in Fig. \ref{fig:cdists_disc} except here we are considering discs extracted as part of an entire system (see Section \ref{sect: multiple discs}). Across the calculations we tend to see characteristic disc radius increase with metallicity.  The median radius of discs in a bound system for the super-solar metallicity calculation is $\sim 65$ au compared to $\sim 20$ au in the lowest metallicity calculation. Around 30\% of discs in the super-solar metallicity calculation have a characteristic radius larger then 100 au compared to around 10\% in the lowest metallicity calculation. This is likely due to the increase of fragmentation due to cooler gas temperatures in the low metallicity calculation (see Section \ref{sect:discussion:opacity}). The characteristic radii tend to be slightly larger for discs in systems than circumstellar discs alone, as to be expected.  We note that both circumstellar discs and discs of systems in the \citet{bate_2012_10.1111/j.1365-2966.2011.19955.x} calculation and the solar metallicity \citet{bate_metallicity_2019MNRAS.484.2341B} calculation have very similar distributions of disc radii.  The circumstellar disc mass distributions are also very similar, although the disc masses of systems are about a factor of two lower in the newer calculation compared to the older calculation.  The two different methods employed by these calculations are not expected to show much difference due to the relatively high densities of the initial molecular cloud. The majority of discs across all the calculations have characteristic radii ranging between 20 and 110 au.

\begin{figure*}
    \centering 
    \includegraphics[scale=0.6]{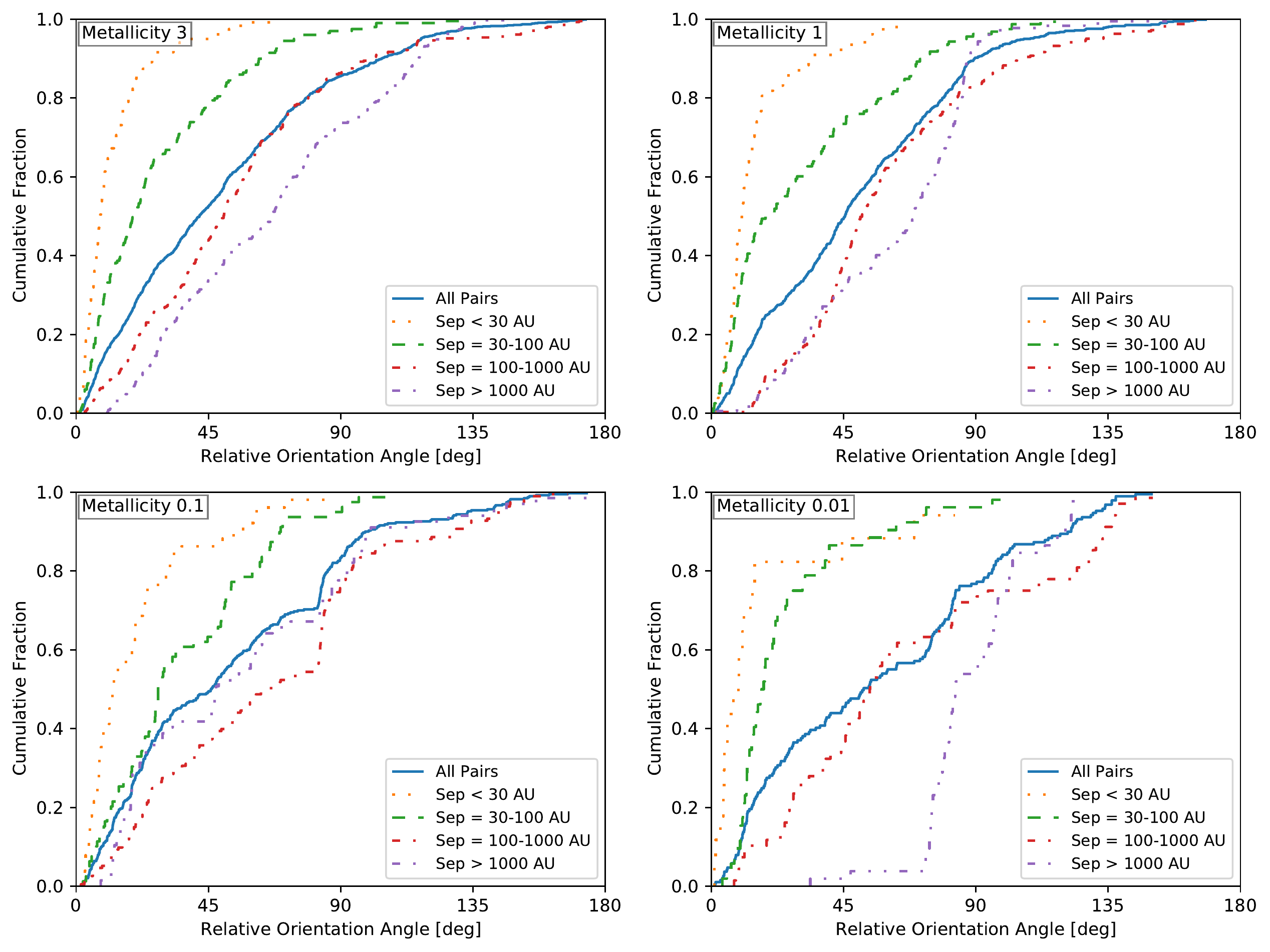}
    \caption{The distributions of relative orientation angle between the two circumstellar discs in bound pairs of protostars. The pairs include binaries, and pairs in triple and quadruple systems. We give the cumulative distributions for four ranges of the semi-major axis: $a<30$ au, $30<a<100$ au, $100<a<1000$ au, and $a>1000$ au. We also plot the cumulative distribution of relative orientation angle for all pairs. As metallicity is decreased the alignment between discs also does, except for close systems ($a<100$ au) at the lowest metallicity ($Z=0.01~{\rm Z_\odot}$). }
    \label{fig:disc_disc}
\end{figure*}

\begin{figure*}
    \centering
    \includegraphics[scale=0.6]{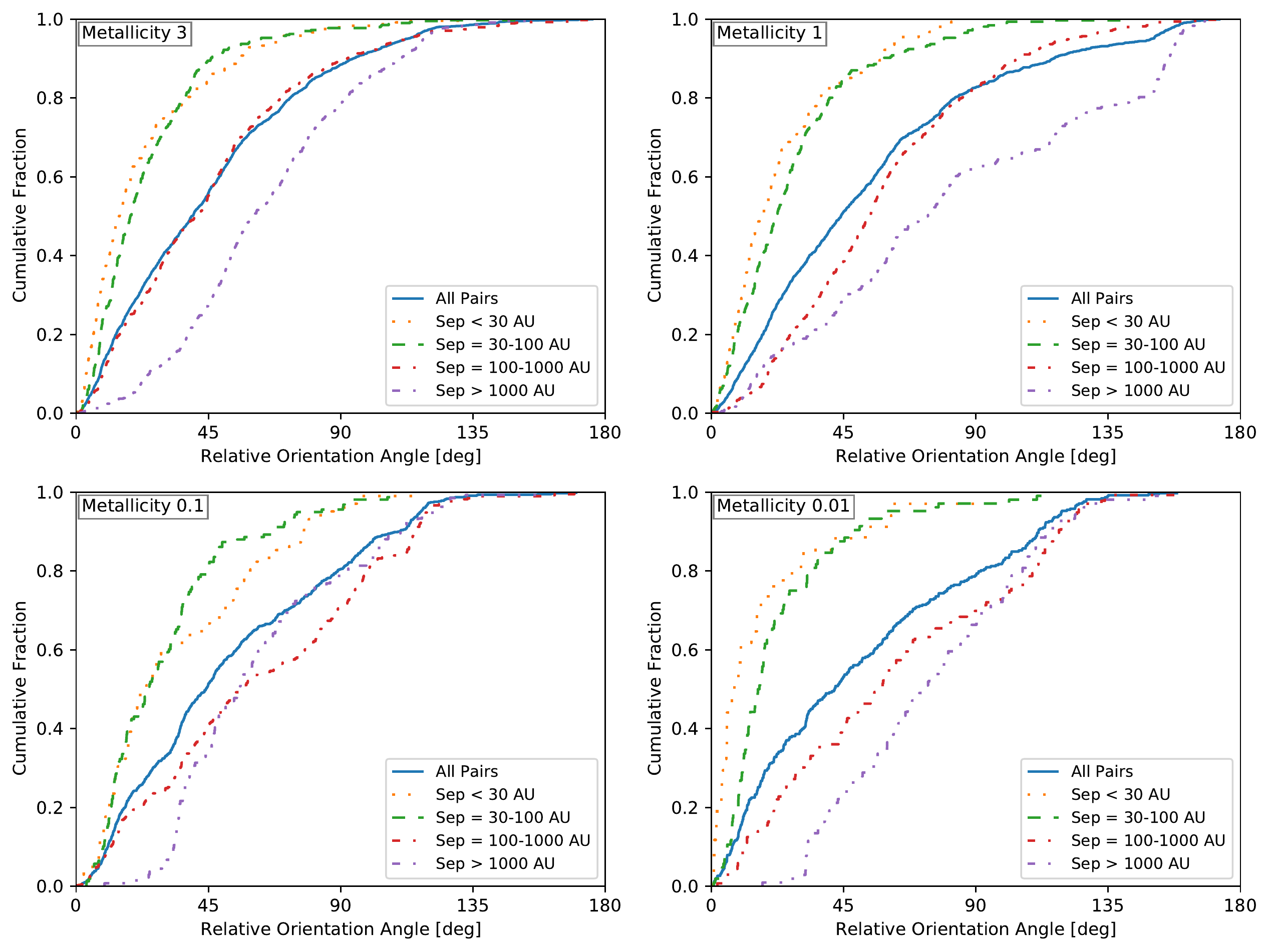}
    \caption{The distributions of the relative orientation angle between each circumstellar disc and the orbital plane of the bound protostellar pairs. The pairs include binaries, and pairs in triple and quadruple systems. We give the cumulative distributions for all pairs, and for those in four ranges of the semi-major axis: $a<30$ au, $30\le a<100$ au, $100\le a<1000$ au, and $a\ge1000$ au. As the metallicity is decreased the distributions for each given separation ranges tend to become flatter, indicating a more random distribution of orientation angles between discs and the orbital plane of the bound pair.  Again, there is an exception for close pairs ($a<100$ au) at the lowest metallicity.  }
    \label{fig:disc_orbit}
\end{figure*}

\begin{figure*}
    \centering
    \includegraphics[scale=0.6]{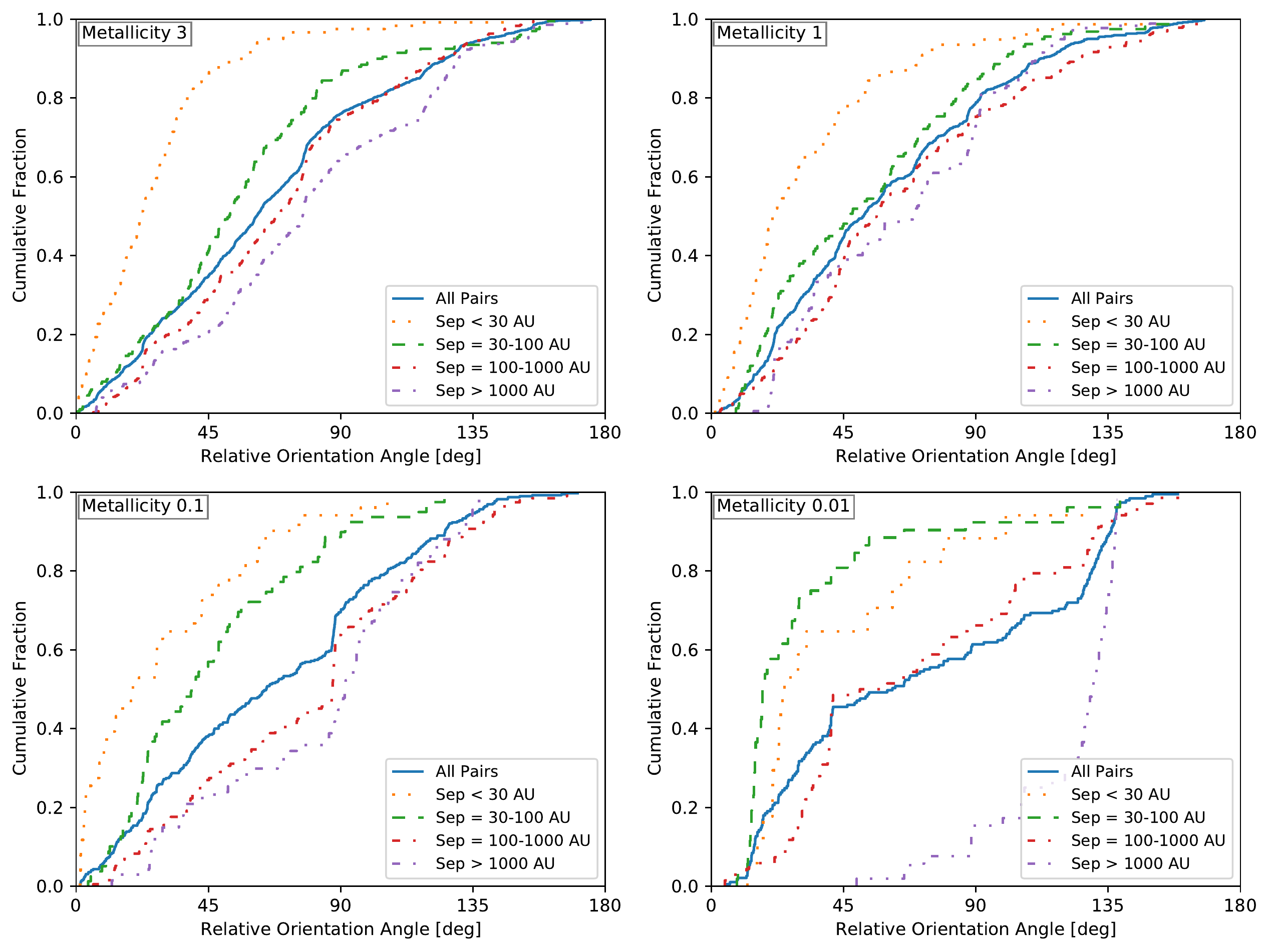}  
    \caption{The distributions of the relative orientation angle between the two sink particle spins of bound protostellar pairs. The pairs include binaries, and pairs in triple and quadruple systems. We give the cumulative distributions for four ranges of the semi-major axis: $a<30$ au, $30\le a<100$ au, $100\le a<1000$ au, and $a\ge1000$ au.  Protostellar spins are generally less well aligned with each other than circumstellar discs are with each other or with the orbit.}
    \label{fig:spin_spin}
\end{figure*}

The right panel of Fig. \ref{fig:cdists_sys} shows the cumulative distributions of the ratio between the total disc mass and the total protostellar mass of the instances of protostellar systems. Overall the distributions tend to be quite flat indicating a wide range of ratios, but that systems in which the total disc mass exceeds the total protostellar mass are very rare even at these young ages.  Again, the lowest metallicity calculation differs slightly from the others in that a greater fraction of systems have low-mass or unresolved discs. 

In Fig. \ref{fig:sys_mass_mass} we plot the cumulative distributions of total system disc mass separated into mass bins of the total system stellar mass ($M<0.1~\text{M}_\odot$, $0.1~\text{M}_\odot<M<0.3~\text{M}_\odot$, and $M>0.3~\text{M}_\odot$). We see there is a clear trend of increasing total disc mass with increasing total stellar mass, except in the lowest metallicity calculation. Additionally we find that higher order systems have a higher total disc mass, although we do not present that data here. We also note that discs about the stars of mass $M>1~\text{M}_\odot$ in the $Z=0.1~{\rm Z}_\odot$ metallicity calculation tend to be more massive that the corresponding discs in the other calculations, although the numbers of systems are relatively small. The distributions of disc masses for the three highest metallicity calculations (apart from the highest mass systems) are quite similar to each other and quite different to the distributions in the lowest metallicity case. In the lowest metallicity case, the disc mass distribution of high-mass ($M>0.3~\text{M}_\odot$) systems is quite similar to that of intermediate mass systems ($0.1~\text{M}_\odot < M < 0.3~\text{M}_\odot$).  In other words, massive discs in massive systems are significantly rarer at $Z=0.01~{\rm Z}_\odot$ than at higher metallicities.  Again, this is likely a result of enhanced cooling at low metallicity and, thus, more frequent disc fragmentation.  Similarly, more than 80\% of discs orbiting very low-mass systems are unresolved, compared to $\approx 60$\% at higher metallicities. 

\subsubsection{Disc orientations of protostellar pairs}

In this section, we investigate how relative orientations of protostellar orbits and spins, and discs in bound protostellar pairs depend on metallicity and orbital separation. We note that distributions of relative orientation angle do not vary much with age and are generally similar to those found by \citet{bate_diversity_of_discs_2018MNRAS.475.5618B}.  For this reason we do not present any graphs of the age dependence here. By bound pairs we mean either a binary system or bound pairs in a higher order system.  A triple system will always contain one pair (due to the hierarchical way in which systems are identified), and a quadruple system may contain either one or two pairs. For discs to be analysed each protostar must have a circumstellar disc. We require each disc to be a mass of $M_d\ge 4.3\times10^{-4}~\text{M}_\odot$ (at least 30 SPH particles) as this is sufficient to determine the angular momentum vector of the disc. For metallicities of $Z=3, 1, 0.1,  0.01~{\rm Z}_\odot$ the numbers of instances of disc pairs are 1076, 735, 390, and 189 respectively.

First, we investigate the relative orientations of the two circumstellar discs in pairs.  In Fig. \ref{fig:disc_disc}, we show the cumulative distributions of the relative orientation angle and its dependence on the separation (semi-major axis) of the protostellar pair for each metallicity calculation. We plot the overall distribution, as well as breaking the samples into four orbital separation ranges: $a<30$ au, $30<a<100$ au, $100<a<1000$ au, and $a>1000$ au. Overall we see that the orientation between discs depends strongly on the separation with a smaller separations leading to a greater degree of alignment between the discs and a larger separation leading to less alignment. In the lowest metallicity case the distribution for all pairs is flatter than for higher metallicity indicating that the disc orientations are starting to become more random. Additionally the differences between the distributions as a function of separation are greatly magnified.  In particular, discs in widely separated ($a>1000$ au) protostellar pairs at the lowest metallicity mostly have relative orientation angles ranging from $70-130\degr$ discs and those with separations $100<a<1000$ have close to a random distribution.  In the lowest metallicity calculation, discs with separations $30<a<100$ au also are more similarly distributed to the discs of separation $a<30$ au than for higher metallicities. We don't see any age dependence on the alignment of discs. \citet{bate_diversity_of_discs_2018MNRAS.475.5618B} found a weak age dependence such that older systems were slighly better aligned. An age dependence is what we might expect due gravitational torques acting on the discs from the bound pair slowly aligning them with the orbital plane, and accretion of gas from outside the system causing discs to align with each other. However, we are only able to study evolution on time scales of $\sim 10^4$ years from these hydrodynamical calculations, and the time scale for significant realignment for most systems is likely much longer.

Next, in Fig. \ref{fig:disc_orbit} we investigate the distributions of relative orientation angle of circumstellar discs and the orbital plane of the bound pairs, plotting orientations in the same four ranges of semi-major axis as in Fig. \ref{fig:disc_disc}. Here we have two values for each bound pair as there are two discs to consider. We find that discs in pairs that are more closely separated than 100 au tend to be well aligned with the orbital plane, although not as well aligned as discs in the closest systems ($a<30$ au) are with each other. As we saw when considering the orientations between discs, the distributions of relative orientation angle for the pairs separated $a>100$ au in lowest metallicity calculation tend to be flatter than the other distributions, with essentially a random distribution of relative angles between zero and $135\degr$. When find no significant dependence of the cumulative distributions of relative orientation angles on age. We may expect older systems to become more aligned as the gravitational torques from the protostars cause the disc to align with the orbital plane, but again we likely do not cover the required time scale to see this effect have a significant impact. Overall, the dependence of the disc-disc relative orientation dependence on separation is somewhat stronger than it is for the alignment of discs with the orbital plane. 

In Fig. \ref{fig:spin_spin} we show the cumulative distributions of the relative orientation angle between the spins of the two sink particles in the bound pairs. The spins are indicative of the angular momenta of the protostar and the inner part of the disc. Overall we see that the spins tend to be less well aligned than the two discs and than discs with the orbital plane of the pair. Spin is only effected by accretion, so if two protostars form in different regions and then become bound we would expect greater misalignment. This is why we see a less overall alignment between spins than we do for discs. In the top right panel we see that there is not a strong dependence on separation in the solar metallicity case, except for the closest ($a<30$ au) pairs which are more well aligned. There is a trend with metallicity such that as the metallicity decreases the spins of sink particles in pairs for which $30<a<100$ au become more aligned, with the angles between spins for sink particles with $a<30$ au and $30<a<100$ au becoming very similarly distributed for low metallicities. In the lowest metallicity calculation, we find a dependence on multiplicity such that spins of pairs in triple and quadruple systems are better aligned than in binary systems. We don't see such a dependence on multiplicity in the three highest metallicity calculations. This could be a signature of an enhanced role of disc fragmentation producing pairs at low metallicity. 

\begin{figure*}
    \centering
    \includegraphics[scale=0.6]{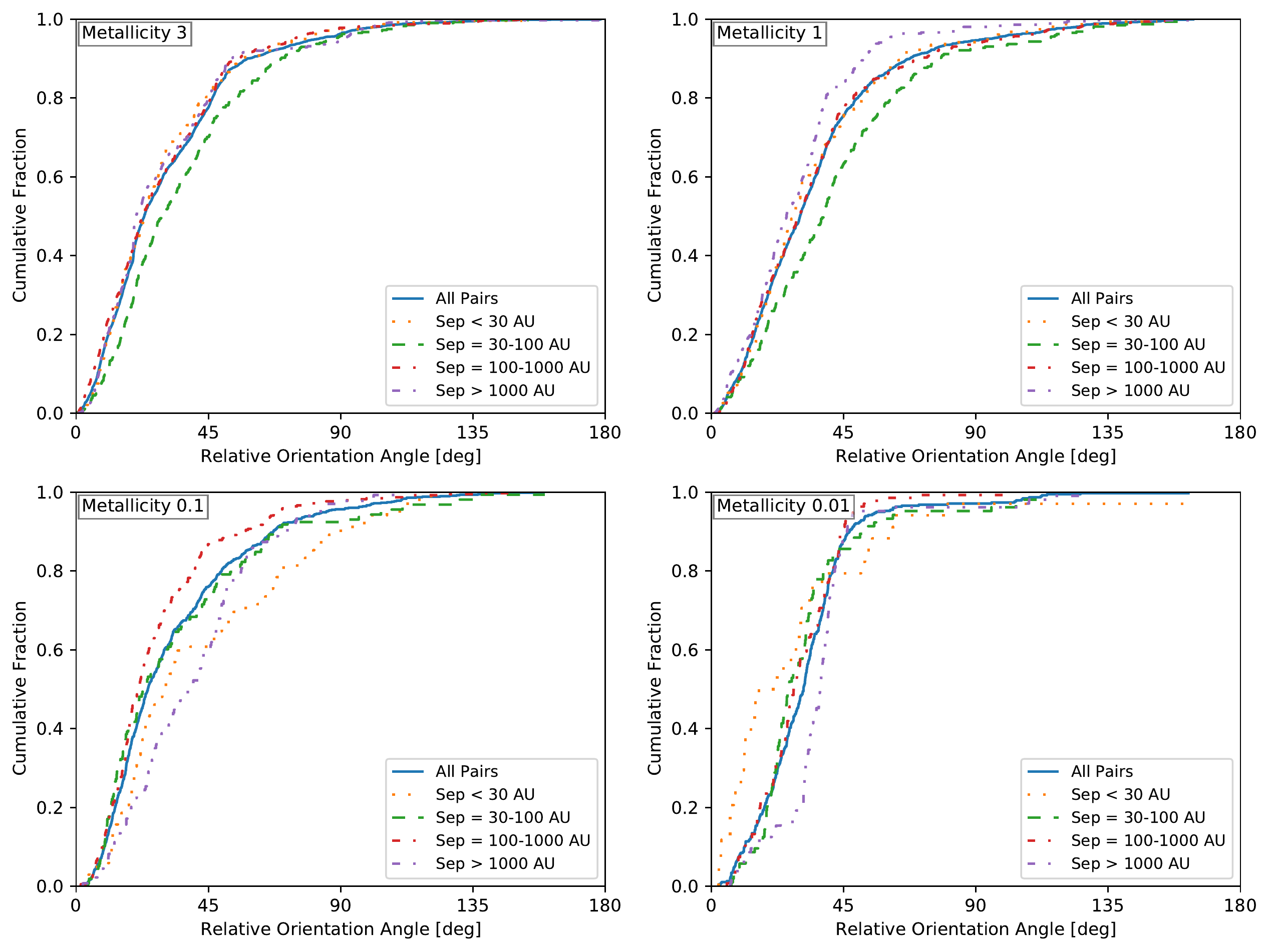} 
    \caption{The distributions of the relative orientation angle between the circumstellar disc of a protostar and the spin of the sink particle (protostar and inner disc) that it orbits, for protostars in bound pairs. The pairs include binaries, and pairs in triple and quadruple systems. We give the cumulative distributions for four ranges of the semi-major axis: $a<30$ au, $30\le a<100$ au, $100\le a <1000$ au, and $a\ge1000$ au. Protostellar discs and spins are generally well aligned with each other, independent of separation of the pair.  There is not much variation with metallicity, but in the lowest metallicity calculation in each semi-major axis separation range $\gtrsim 80\%$ of discs and spins are aligned to within $45\degr$, slightly more aligned than the higher metallicity cases.}
    \label{fig:disc_spin}
\end{figure*}

Finally, we compare the relative orientation angles between the angular momentum vector of the disc and the sink particle spin in Fig. \ref{fig:disc_spin}. Across all four calculations we see a strong preference for alignment. This is to be expected as the spin of a sink particle represents the angular momentum of the protostar and inner part of the disc which accretes from the outer disc. We see that there is no significant dependence of relative orientation distribution between discs and sink particle spin on separation. Each protostar and its disc evolves in a similar manner with accretion and/or gravitational torques reorientating the outer disc and the sink particle spin continually trying to `catch up' by accreting from the outer disc (see the end of Section 5.3.2 of \citealt{bate_diversity_of_discs_2018MNRAS.475.5618B} for further discussion).  With the lowest metallicity we see a slightly better alignment between discs and sink particle spins. As we have seen above, a greater fraction of the instances of protostars tend to have low mass (or unresolved) discs in the lowest metallicity calculation.  It may be that many of these instances of discs have not had much recent accretion and, therefore, the sink particle spins and outer discs have had time to evolve to become more closely aligned.  We don't detect any dependence on age or multiplicity for the relative orientation angles between discs and protostellar spins. 

\section{Discussion}\label{sect:discussion}

\subsection{Effect of opacity}\label{sect:discussion:opacity}

 As the metallicity is decreased in the calculations, the opacity is lowered which leads to more rapid cooling of high-density gas (due to reduced optical depths) and fragmentation becomes more likely to occur.  Therefore, any given disc should be more gravitationally unstable. When examining the close binary fraction of solar type stars ($M_1 = 0.6 - 1.5~\text{M}_\odot$) \citet{Moe2018_close_binary} find it to be strongly anticorrelated to metallicity. They suggest enhanced disc cooling due to reduced opacity causes an increased rate of disc fragmentation leading to a higher rate of close binaries. 
 
\citet{bate_metallicity_2019MNRAS.484.2341B} obtained a similar anticorrelation between the close binary frequency and metallicity in his radiation hydrodynamical calculations.  However, when he investigated the cause of the enhanced close binary fraction at low metallicities, he found that classical disc fragmentation (i.e. a circumstellar disc fragmenting into one or more objects which then form a close binary with the original protostar) is not the main cause of the of the anticorrelation.  Rather it arises due to enhanced  fragmentation on small scales in general -- both small-scale fragmentation within collapsing molecular cloud cores and disc fragmentation contribute. 

This higher rate of fragmentation in the lower metallicity calculations is the reason for the differences in disc properties that we have found with different metallicities.  The anticorrelation of close binary frequency with metallicity means that more circumstellar discs are disrupted or truncated at lower metallicities and are unresolved or have very low masses (e.g., Fig \ref{fig:cdists_disc}).  Similarly, those discs that are resolved tend to be smaller with low metallicities. Conversely, a greater fraction of protostars have resolved discs as the metallicity is increased, and those discs tend to be larger. 

\begin{figure}
    \centering
    \includegraphics[scale=0.6]{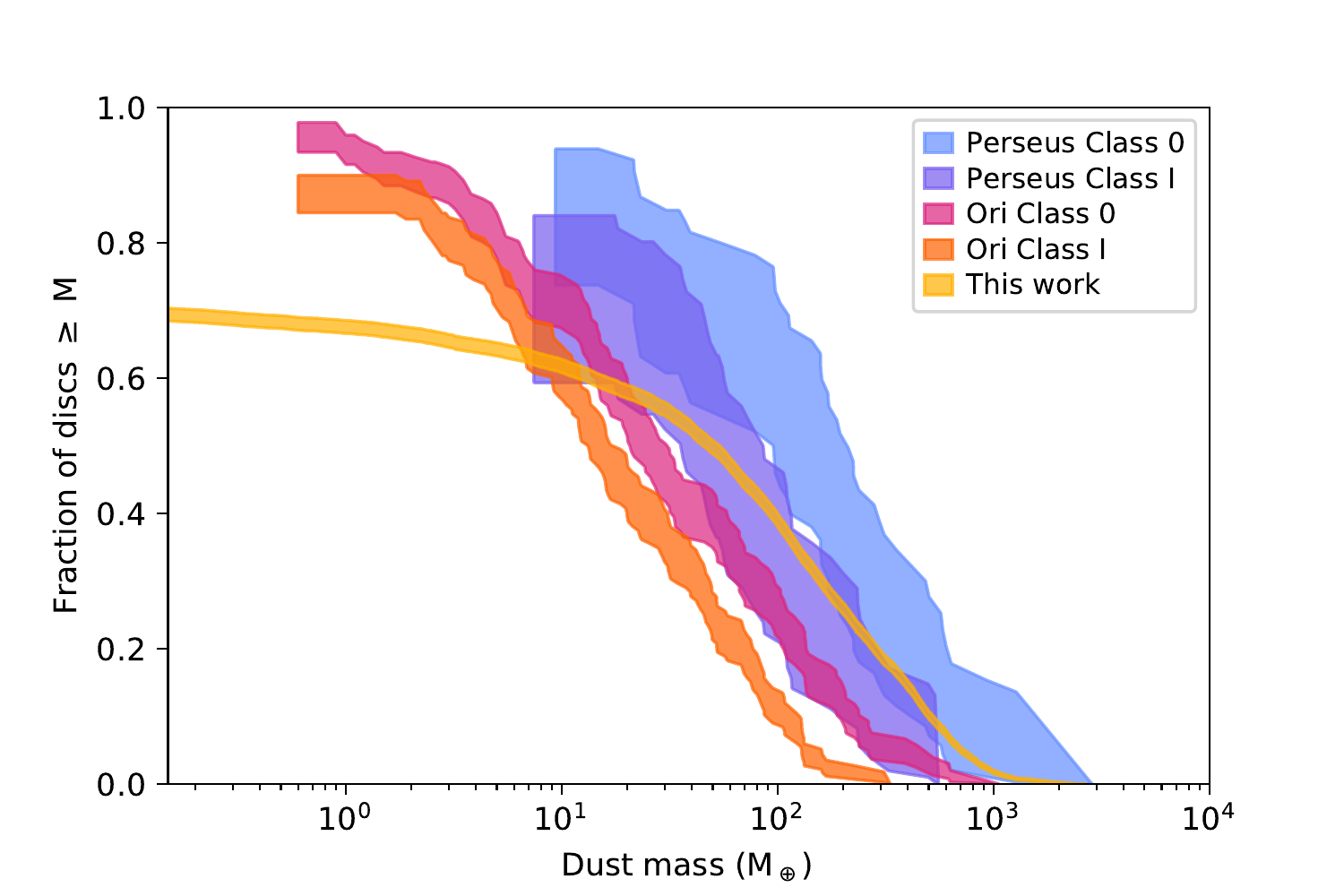} 
    \caption{Dust mass distributions of discs around protostars in protostellar systems from the solar metallicity calculation of \citet{bate_metallicity_2019MNRAS.484.2341B} and for discs of Class 0 and I objects taken from surveys of nearby star-forming regions. We use disc masses from Perseus \citep{tychoniec_2020_vandam} and Orion \citep{tobin_vandam_2020ApJ...890..130T}. We make use of the Kaplan-Meier estimate for left censored data as implemented in the Python package \textit{lifelines} \citep{cameron_davidson_pilon_2021_4816284}. The shaded regions for the Orion discs and the simulation indicate $1\sigma$ confidence intervals, and a confidence intervals of $3\sigma$ for Perseus as per \citet{tychoniec_2020_vandam}.}
    \label{fig:disc_dust_mass_class_0}
\end{figure}

\subsection{Some limitations}

As mentioned in Section \ref{sec: sink particles}, although the hydrodynamical calculations analysed in this paper include radiative transfer, they do not include stellar radiative feedback from the sink particles that model the young stars themselves \citep[c.f.][]{Offneretal2009,Urban2010,JonBat2018b,MatFed2020}.  Radiation generated within the sink particle radius (0.5 au) is neglected.  Although the radiative feedback for very young, low-mass protostars is relatively weak, gas temperatures near protostars are likely to be underestimated \citep[see][for a detailed discussion]{bate_2012_10.1111/j.1365-2966.2011.19955.x}. Thus, disc fragmentation may be expected to be somewhat more prevalent in the calculations analysed here than if stellar radiative feedback were present \citep[e.g.][]{offner2010formation,MatFed2021}. Despite this, we expect that the trends with metallicity seen in both \citet{bate_metallicity_2019MNRAS.484.2341B} and this paper should persist if stellar radiative feedback were included, although the quantitative numbers may change slightly.  The reason is that the increased small-scale fragmentation with decreasing metallicity due to lower opacities that was identified by \citet{bate_metallicity_2019MNRAS.484.2341B} should persist since although some of this fragmentation is due to disc fragmentation (where existing protostars are present and where radiative feedback may affect the outcome), much of the small-scale fragmentation that leads to multiple system formation occurs within pre-stellar molecular cloud cores (where there is no protostar at the time of fragmentation and, thus, stellar radiative feedback is unimportant).  Since this enhanced small-scale fragmentation is responsible for both the enhanced close binary frequency at lower metallicities found by \citet{bate_metallicity_2019MNRAS.484.2341B}, and the smaller disc radii and slightly lower system disc masses that we find here, we expect these trends would persist if stellar feedback were included.

Although the thermodynamical evolution of the calculations treats dust and gas temperatures, the calculations do not include dust dynamics or grain growth.  Although dust grain migration in discs may be expected to be relatively weak for the young protostars studied here, it should be noted that dust grain growth and, therefore, dynamics should also be metallicity dependent.  In particular, at lower metallicities, dust grains may be expected to grow more slowly due to their decreased grain number density and, therefore, even if the gas distribution of a low-metallicity disc and a high-metallicity disc were identical, the dust opacity and dust migration in the discs is likely to differ. This shouldn't have any significant effect on the gas mass and radii distributions obtained in this paper, but since observations usually use dust emission to measure the masses and sizes of discs there may be an additional effect of metallicity on the apparent dust masses and disc sizes due to the dependence of dust evolution on metallicity that should be noted by observers.

\subsection{Comparison with previous theoretical results}

The trends we find in circumstellar disc mass and size are in general agreement with those found by \citet{bate_diversity_of_discs_2018MNRAS.475.5618B}, and the statistical properties of discs for the solar metallicity calculation in particular are in very close agreement with those obtained by \citet{bate_diversity_of_discs_2018MNRAS.475.5618B}. This isn't too surprising even though the calculations we consider include thermodynamical effects not included in the previous population study of disc properties as the earlier approach is a good approximation at the high densities of the molecular clouds considered in both studies. Similarly, we also find that discs with no encounters are generally more massive and larger than discs that have had interactions. The relative alignments in bound pairs of protostars between discs, the orbital plane, and sink particle spins also have similar dependencies on separation, although unlike \citet{bate_diversity_of_discs_2018MNRAS.475.5618B} we do not detect a dependence on age of the systems.

The new aspect of our study is that we consider whether and how the statistical properties of discs depend on metallicity.  This is the first time such a statistical analysis of disc properties at different metallicities has been carried out based on the results of radiation hydrodynamical simulations.  

However, studies of how the evolution of individual molecular cloud cores and discs are affected by their metallicity have been carried out before.  For example,  \citet{Machida2008, Machidaetal2009, TanOmu2014, bate_metallicity_2014MNRAS.442..285B} all showed that fragmentation increases with lower metallicity.  Recently, \citet{vorobyov_low_metal_accretion_2020A&A...641A..72V} studied the early evolution of individual protostellar discs in simulations for metallicities ranging $Z=1-0.01~\text{Z}_\odot$.  They focus on the gravitational instability of discs and periodic accretion bursts particularly in low metallicity discs. The accretion rates in the low metallicity discs during the embedded disc phase (40 and 320 kyr for the $Z=0.01~\text{Z}_\odot$ and $Z=0.1~\text{Z}_\odot$ simulations respectively) are higher in the $Z=0.01~\text{Z}_\odot$ simulation. In general agreement with this work, we note that the mean accretion rates of discs at the end of each of the star cluster formation simulations increase as metallicity decreases \citep[][see Table 3]{bate_metallicity_2019MNRAS.484.2341B}.  The discs in the cluster simulations are at a similar age to those simulated by \citet{vorobyov_low_metal_accretion_2020A&A...641A..72V}.

\subsection{Comparison with Class 0/I objects}

The calculations come with a caveat that the discs that are analysed aren't very old (the oldest is $\sim 10^5 \text{yr}$). Even though objects don't have a well defined age sequence \citep{Kurosawaetal2004,Offner_etal2012}, protostars of this age are generally thought be Class 0 objects. While a Class 0 protostar is defined by observational signatures that indicate the presence of a substantial envelope \citep{andre_class_0_1993}, there is now strong evidence that discs can and do grow at this stage \citep[e.g.][]{yen_class_0_0I_2015,tobin_vandam_2020ApJ...890..130T}. The catalogue of discs around Class 0 objects is growing with the advent of the Atacama Large Millimeter/submillimeter Array (ALMA). For example, ALMA confirmed the earlier result \citep{tobin_2012_l1527irs} of a Class 0 object with a rotationally supported disc around L1527 IRS with radius \citep{ohashi_2014_l1527irs,sakai_2014_l1527,aso_2017_l1527irs}.  Currently, the largest observational samples of Class 0 and Class I discs are those from the VLA/ALMA Nascent Disk and Multiplicity (VANDAM) survey of Perseus protostars \citep{tychoniec_2018_vandam,tychoniec_2020_vandam} and the VANDAM survey of  Orion protostars \citep{tobin_vandam_2020ApJ...890..130T}.  The work of \citet{tychoniec_2020_vandam} aims to provide more accurate disc mass determinations than the earlier work of \citet{tychoniec_2018_vandam} by considering the effects of large dust grains on the opacities. Therefore, we use the results of the more recent study when making comparisons with the simulations.

\subsubsection{Disc dust masses}

In Fig. \ref{fig:disc_dust_mass_class_0} we plot the cumulative distributions of the disc dust masses from VANDAM surveys of Orion and Perseus Class 0/I protostars, along with the equivalent distribution for instances of discs of protostellar systems from the solar metallicity calculation.  We use the Kaplan-Meier estimator, as implemented in the Python package \textit{lifelines} \citep{cameron_davidson_pilon_2021_4816284}, with left censoring to account for upper limits on the observed disc masses. We provide shaded  $1\sigma$ confidence intervals for discs in Orion and the simulations, and $3\sigma$ for discs in Perseus \citep{tychoniec_2020_vandam}.  Dust masses for the simulated discs are determined by using the standard dust to gas ratio of 1:100. The shaded area for the simulated discs is narrow due to the large number of instances of discs that are identified.  The Kaplan-Meier estimator is used here as it becoming a standard tool for observational studies of discs. 

The form of the mass distribution of simulated discs is similar to the observed distributions.  The flattening of the mass distribution of the simulated discs below $\approx 50~\text{M}_\oplus$ is due to the limited resolution of the calculation.  Discs that are poorly resolved viscously evolve quicker than they should and, thus, have lower masses than they would at higher resolution \citep[see][]{bate_metallicity_2019MNRAS.484.2341B}.  Above this mass, the simulated discs are nicely bracketed between the mass distributions of the Perseus and Orion Class 0 objects.  The Perseus Class 0 disc dust masses are typically about a factor of two higher than for the simulated discs, and the Orion Class 0 masses are about a factor of two lower.  The observed Class I discs have lower masses than the Class 0 discs by factors of 2--3, consistent with them typically being more evolved.
The mean disc dust mass for the Class 0 and Class I in Orion are $25.9^{+7.7}_{-4.0}$ and $14.9^{+3.8}_{-2.2}~\text{M}_\oplus$ respectively and the median disc dust mass for Class 0 and Class I in Perseus are $158$ and $52$ M$_\oplus$ respectively. The median mass of system discs in the solar metallicity calculation is $49~\text{M}_\oplus$.  Given that there is considerable uncertainty in determining the masses of observed discs \citep[e.g.,][]{tychoniec_2020_vandam}, the simulated disc masses are in good agreement with the observed Class 0 disc mass distributions.






\subsubsection{Disc Radii}

\begin{figure}
    \centering
    \includegraphics[scale=0.6]{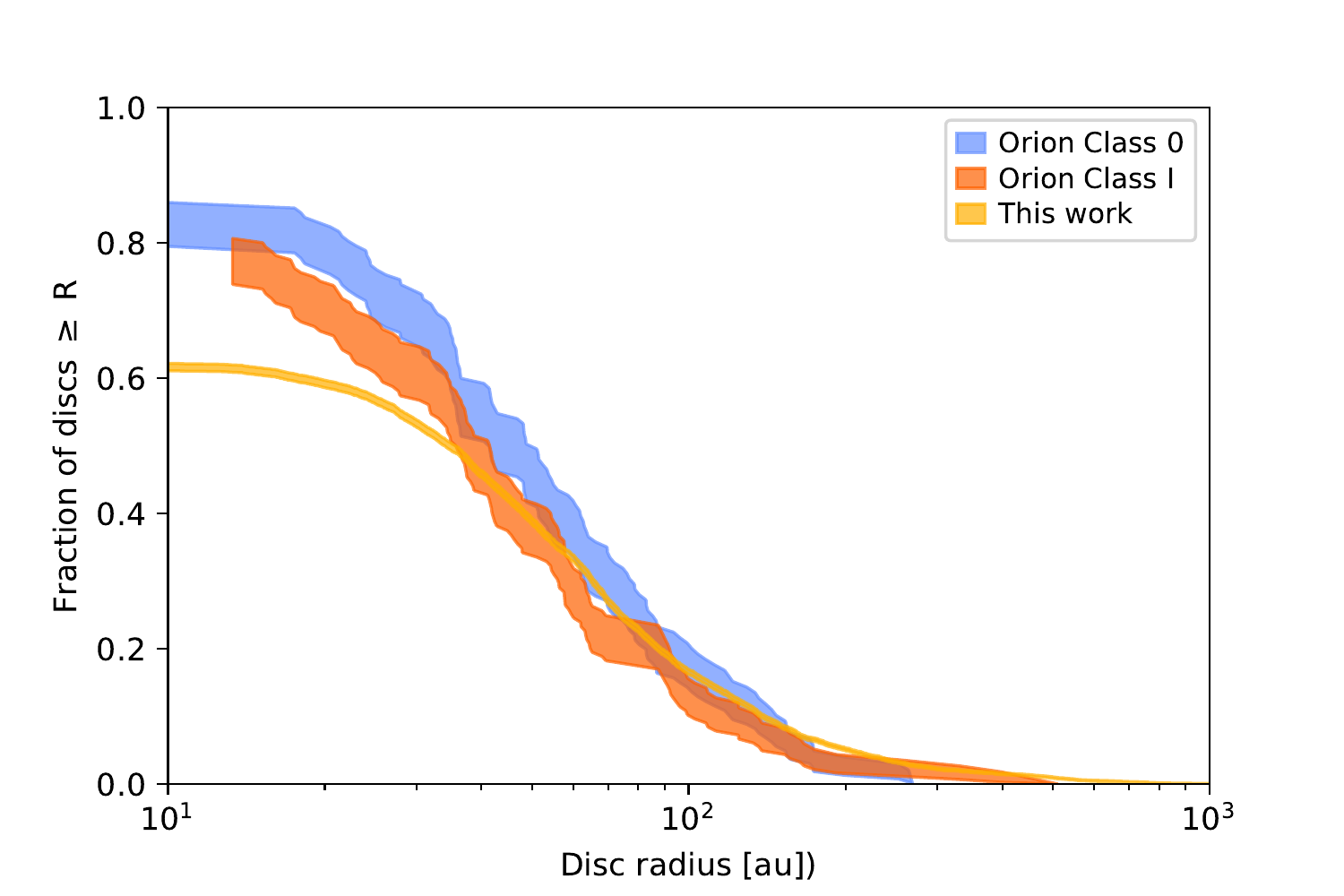}
    \caption{Disc characteristic radii distributions from protostellar systems from the solar metallicity calculation of \citet{bate_metallicity_2019MNRAS.484.2341B} and for the discs of Class 0 and Class I objects in Orion taken from the VANDAM survey \citep{tobin_vandam_2020ApJ...890..130T}. We use the Kaplan-Meier estimate for left censored data where the width shaded region is the confidence interval at $1\sigma$ (68\%). Observed discs with radii $\le10$au and non-detections are treated as upper limits.}
    \label{fig:class I rad}
\end{figure}

In Fig. \ref{fig:class I rad} we plot the cumulative distributions of Class 0/I disc radii from the VANDAM survey of Orion protostars \citep{tobin_vandam_2020ApJ...890..130T}, and the equivalent distribution for systems from the solar metallicity calculation. We use the data of all discs from the 0.87mm ALMA observations as at this wavelength the derived radius of a given disc is assumed to be close to the gas radius of the disc. For the observed discs, we treat discs with radii $\le 10$ au and non-detections as the upper limits in the analysis. 

Immediately it is clear that the radii of discs from the calculation have a remarkably similar distribution to those observed discs for radii $r_\text{c} \gtrsim 40$~au. Again, the turnover for the simulated discs at smaller radii is due to the limited numerical resolution of the simulations.  

\subsection{Comparison with Class II objects}

Over the past few years, improvements in (sub-)millimetre resolution have given us great ability to conduct large surveys of protoplanetary discs in nearby star-forming regions. Whilst Class II objects are more evolved than any of the protostars we consider in this paper, it is still useful to compare the empirical trends found from observational studies and those we present. 

\subsubsection{Disc dust mass}

Here we compare disc dust masses derived from surveys of various star forming regions with the masses of the simulated discs from the solar metallicity calculation of \citet{bate_metallicity_2019MNRAS.484.2341B}.  Again we use the standard dust to gas ratio of 1:100 to convert the gas masses of the simulated discs into dust masses.

We compare the disc dust masses we obtained from our solar metallicity calculation with those derived from observational surveys of Class II objects. We use surveys of: the Lupus star forming region \citep{ansdell_2016_lupus}, the Orion Nebula Cluster (ONC) \citep{eisner_2018_onc}, the Orion Molecular Cloud-2 (OMC-2) \citep{van_terwisga_2019_omc2}, $\rho$ Ophiuchus \citep{cieza_2019_rho_oph}, and $\lambda$ Orionis \citep{ansdell_2020_lambda_ori}. Each derived disc dust mass takes the dust temperature to be $T_\text{dust}=20~\text{K}$. We note that for the OMC-2 observations we haven't included non-detections, hence the fraction of discs with mass $\ge M_{\text{dust}}$ begins at 1.

\begin{figure}
    \centering
    \includegraphics[scale=0.6]{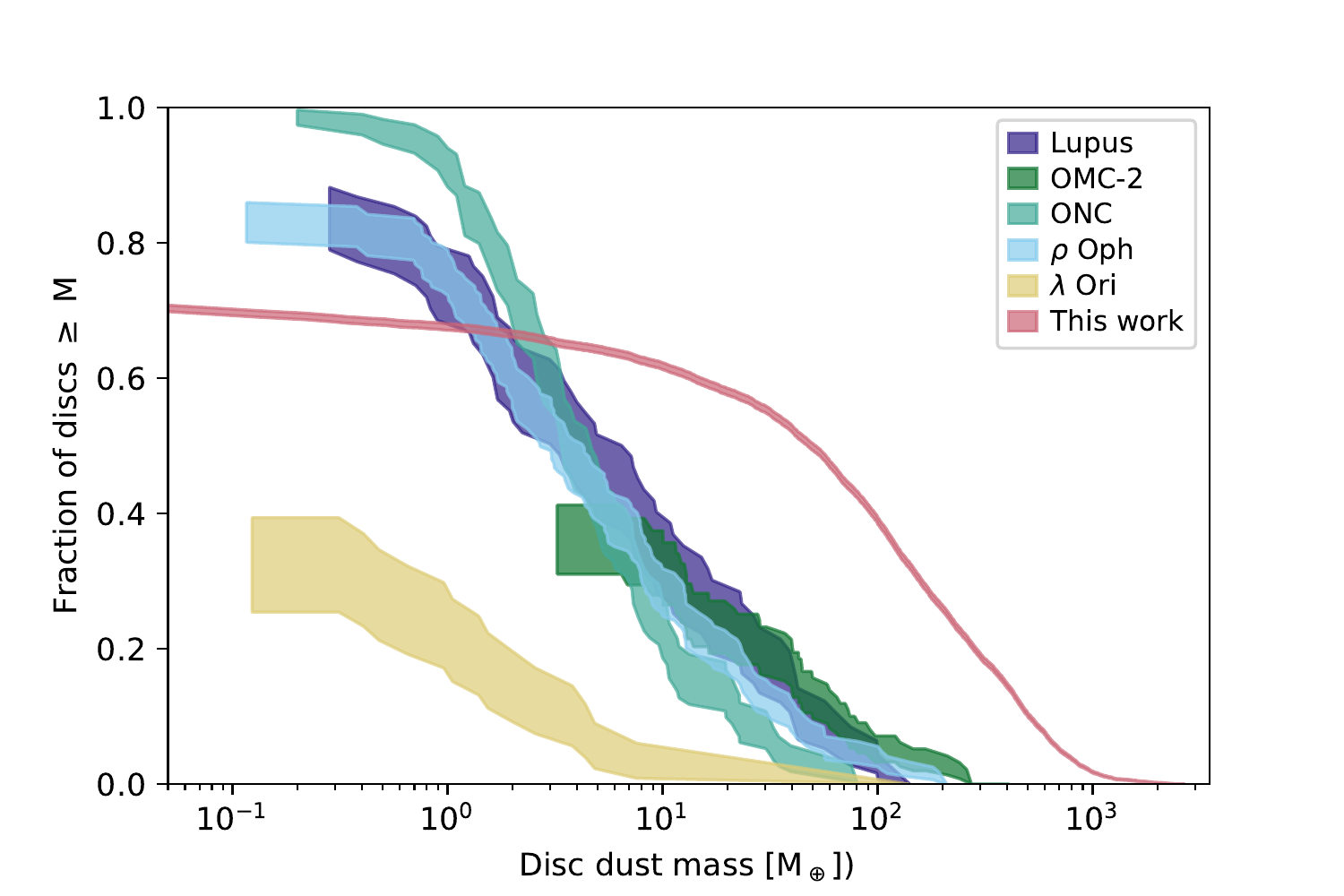}
    \caption{Dust mass distributions of discs of protostellar systems from the solar metallicity calculation of \citet{bate_metallicity_2019MNRAS.484.2341B} and of the discs of Class II objects taken from surveys of different star forming regions. We use surveys of Lupus \citep{ansdell_2016_lupus}, the Orion Nebula Cluster \citep{eisner_2018_onc}, $\rho$ Ophiuchus \citep{cieza_2019_rho_oph}, the Orion Molecular Cloud-2 \citep{van_terwisga_2019_omc2}, and $\lambda$ Orionis \citep{ansdell_2020_lambda_ori}. We make use of the Kaplan-Meier estimate for left censored data as implemented in the Python package \textit{lifelines} \citep{cameron_davidson_pilon_2021_4816284}. The shaded regions indicate a $1\sigma$ ($\approx 68\%$) confidence intervals.  The observed disc masses of Class II objects tend to be much lower than the simulated discs, as is to be expected if the Class II objects are much more evolved. } 
    \label{fig:class_ii_dust}
\end{figure}

In Fig. \ref{fig:class_ii_dust} we plot the cumulative distributions of disc dust masses from the above surveys and the solar metallicity calculation. We use the Kaplan-Meier estimator with left censoring to account for the upper limits on the observed disc masses. An observation that is deemed to be a non-detection is censored. We plot shaded $1\sigma$ ($\approx 68\%$) confidence intervals. 

The mean dust masses for Lupus, ONC, OMC-2, $\rho$ Ophiuchus, and $\lambda$ Orionis are $15\pm3~\text{M}_\oplus$, $8\pm1~\text{M}_\oplus$, $67\pm9~\text{M}_\oplus$, $19\pm4~\text{M}_\oplus$, and $12\pm0.1~\text{M}_\oplus$ respectively. The mean dust mass of discs of systems from the solar metallicity calculation is $165~\text{M}_\oplus$.
It is not surprising that disc dust masses from the simulation are considerably higher than those observed as they are much younger. The most massive discs in the simulation are $\sim 1$ order of magnitude more massive than the most massive discs in the Lupus, ONC, $\rho$ Ophiuchus and OMC-2 regions, and around 2 orders of magnitude more massive than the most massive discs in $\lambda$ Orionis. 

\subsubsection{Disc radii}

Here we compare the statistical properties of disc radii in systems formed in the solar metallicity calculation (see Section \ref{sect: multiple discs}) with the radii derived from observations of Class II objects. We use radii of discs in the ONC \citep{eisner_2018_onc}, Taurus and Ophiuchus \citep{tripathi_2017_tau_ophi}, Lupus \citep{2017A&A...606A..88T}, Upper Scorpius OB association (Upp Sco) \citep{barenfeld_2017_upp_sco} regions. Note the survey of Taurus and Ophiuchus \citep{tripathi_2017_tau_ophi} includes 50 discs from the Taurus and Ophiuchus regions, and 9 discs from other regions. Each of these surveys base the disc radii on dust continuum observations. The radii for the ONC, Taurus and Ophiuchus are computed using a Gaussian half width at half maximum whereas Lupus and Upp Sco inferred the radii are from the exponential cutoff radius of a power-law disc \citep[see][]{2017A&A...606A..88T}. 

\begin{figure}
    \centering
    \includegraphics[scale=0.6]{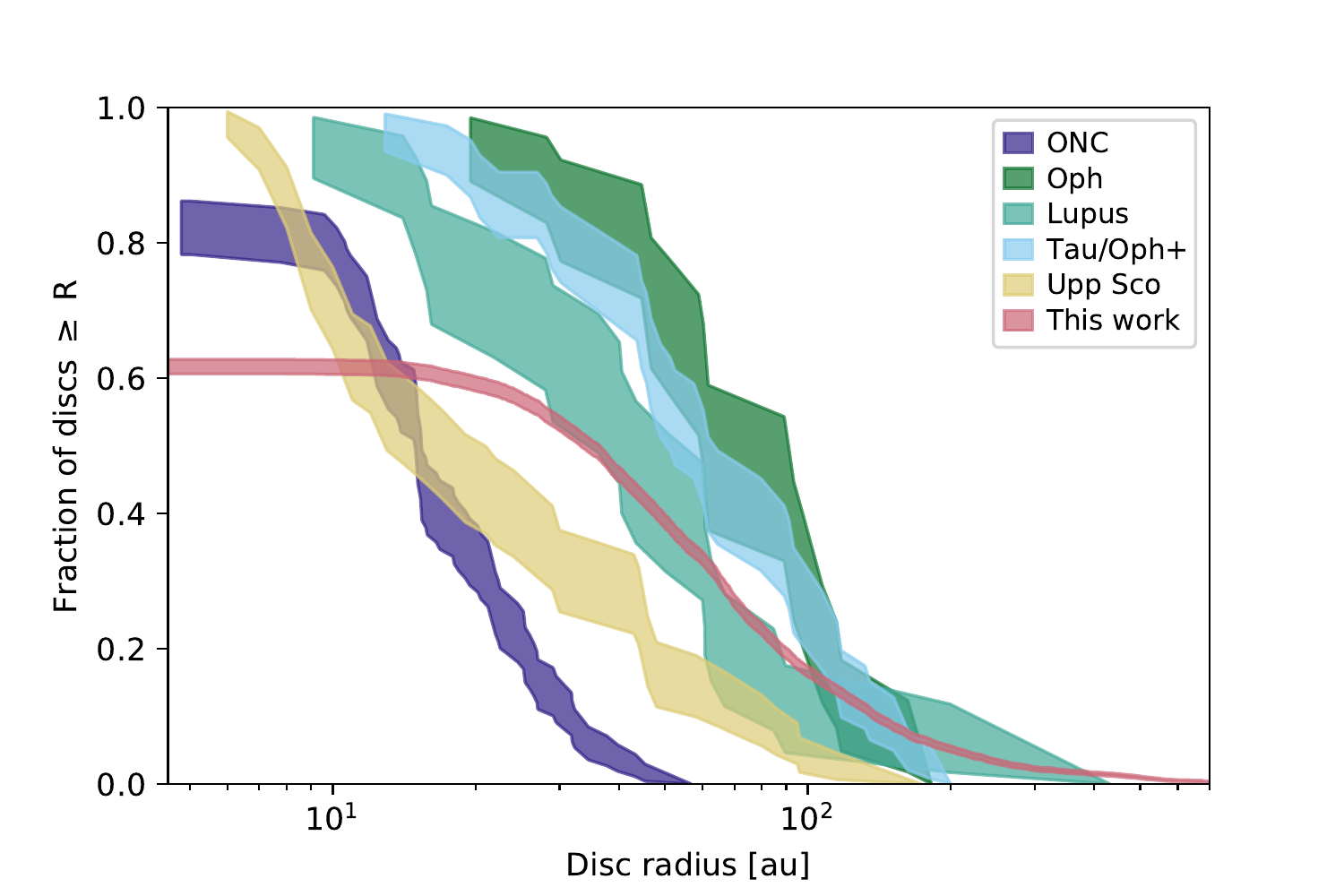}
    \caption{Cumulative distributions of the characteristic radii of discs of systems from the solar metallicity calculation of \citet{bate_metallicity_2019MNRAS.484.2341B}, and discs observed in the Orion Nebula Cluster \citep{eisner_2018_onc}, Ophiuchus \citep{tripathi_2017_tau_ophi}, Taurus and Ophiuchus \citep{tripathi_2017_tau_ophi}, Lupus \citep{2017A&A...606A..88T}, and Upper Scorpius OB association \citep{barenfeld_2017_upp_sco} regions. We use the Kaplan-Meier estimator to plot the cumulative distributions in order to take into account upper limits (unresolved discs) with the shaded region showing a $1\sigma$ ($\simeq 68\%$) confidence interval.}
    \label{fig:rad_class_II}
\end{figure}

In Fig. \ref{fig:rad_class_II} we plot the cumulative distributions of disc sizes taken from the observational surveys, and of characteristic disc radii of systems from the solar metallicity calculation. Again, we use the Kaplan-Meier estimate with $1\sigma$ ($\simeq$68\%) confidence intervals. As we have done with the observed discs we left censor simulated discs that are unresolved (i.e. any disc with a gas mass $<0.01\text{M}_\odot$).  

Immediately we note that there is a large dispersion in the observed disc radius distributions from region to region.  This dispersion may arise due to the different ages of the regions, different evolutionary processes (e.g. photoevaporation), and/or different initial conditions.  The distribution of disc radii from the simulation is slightly less steep than that of the ONC, Lupus, and Taurus and Ophiuchus.  The discs in Lupus have a similar radius distribution to the simulated discs. The discs in the ONC are noticably smaller than in the other regions.  The ONC has a relatively higher stellar density than, for example, Lupus and Ophiuchus, which may explain the high fraction of compact discs and lack of large discs.  However, the initial molecular cloud density for the solar metallicity calculation is also quite high ($n_\text{H} = 6 \times 10^4$~cm$^{-3}$) and this does not lead to unusually small discs.  The discs in the ONC are also subject to intense radiation from the massive stars so photoevaporation may also be the cause of its smaller discs.

\subsection{Disc alignments}

Observationally, many examples of binary protostellar systems with circumstellar discs that are misaligned with each other have been found over the past couple of decades (see Section 3.2 of \citealt{bate_diversity_of_discs_2018MNRAS.475.5618B} for a recent summary).  Early evidence for systems where discs were presumably misaligned with binary orbits when the systems were young came from the observed misalignment of stellar spins and binary orbital planes \citep{Weis1974,Guthrie1985}. \cite{Hale1994} found a preference for alignment for binary separations $a \lesssim 30$~au with random uncorrelated stellar rotation and orbital axes for wide systems.  Similar dependence of the alignment of protostellar (sink particle) spins on the orbital separation of binaries has been reproduced by past hydrodynamical simulations of star cluster formation \citep[e.g.][]{bate_2012_10.1111/j.1365-2966.2011.19955.x,bate_metallicity_2014MNRAS.442..285B}

Recently, \citet{masataka_2020_alignment} examined five star forming regions (Lupus, $\rho$ Ophiuchi, Orion, Taurus, Upper Scorpius) to see whether or not discs were aligned with each other on larger scales than bound stellar systems. They find no overall alignment of discs in all regions except in a sub-region of Lupus (Lupus III), where they find 16 discs that differ from random alignment at the $2\sigma$ level. These discs appear to be aligned with the filamentary structure of the cloud. Carrying out a similar analysis of discs formed in the calculations analysed here, we find no overall alignment of discs. We expect to find no overall alignment of discs in the calculations as the initial cloud has little net rotation and we do not include magnetic fields.

\subsection{Discs at low metallicity}

The disc fraction in young low metallicity star clusters is observed to be lower than in clusters of higher metallicity at a similar age ($\sim 0.5~\text{Myr}$), providing evidence that discs have a shorter lifetime at low metallicities \citep{yasui_low_metal_disc_lifetime_2009,yasui_2016_low_metal_sh_2_208, Yasui_et_al2021}.  \citet{ercolano_metallicity_planet_formation_2010} show that these observations may be due to the metallicity dependence of disc dispersal by photoevaporation.  We are not able to follow discs in the hydrodynamical calculations to these ages (the oldest discs in the lowest metallicity calculation are $\sim 10^4$ yr old), and photoevaporation is not included in the calculations. However, we do find that young discs at lower metallicity have smaller radii and a higher fraction of systems have unresolved or low-mass discs (Fig. \ref{fig:cdists_sys}), which may also help to explain the observed metallicity dependence of disc fraction in young clusters.

\subsection{Planet formation}

The occurrence of giant planets orbiting solar type stars has been shown to decrease with decreasing metallicity \citep{Gonzalez1997, Santos_etal2001,FisVal2005,johnson_giant_planets_2010}, and the dust mass of debris disc stars with planets correlate with stellar metallicity \citep{chavero_trends_in_metallicity_2019}.  The core accretion mechanism for giant planet formation has been shown to be less efficient at producing giant planets in metal-poor discs \citep[e.g.,][]{IdaLin2004}, and \citet{matsumura_2021_n_body_sims_ii} find that in low mass and/or low metallicity discs the formation of giant planets is difficult, and planet formation overall tends to be slower.  More rapid disc dispersal at lower metallicities would also inhibit giant planet formation \citep{ercolano_metallicity_planet_formation_2010}.  
If the initial disc properties also depend on metallicity, as we find here, with metal-poor discs having smaller radii and a higher fraction of systems have unresolved or low-mass discs (Fig. \ref{fig:cdists_sys}), this will also contribute to the paucity of massive planetary systems around metal-poor stars.

\section{Conclusions} \label{sect:concs}

We have presented an analysis of the protostellar discs produced in the four radiation hydrodynamical calculations of star cluster formation with metallicities varying from 1/100 to 3 times solar metallicity first published in \citet{bate_metallicity_2019MNRAS.484.2341B}. These calculations are have identical initial conditions except for their metallicity and, thus, their initial gas and dust temperatures.  Our analysis closely follows the analysis of the statistical properties of discs in an earlier solar metallicity calculation \citet{bate_diversity_of_discs_2018MNRAS.475.5618B}.  We obtain very similar statistical properties from the more recent solar metallicity calculation as those that were obtained by \citet{bate_diversity_of_discs_2018MNRAS.475.5618B}, but we are able to explore the metallicity dependence of disc properties.

We have the following conclusions:

\begin{enumerate}
    \item Higher metallicity typically leads to the formation of larger discs. This is the case for discs of individual protostars, discs of protostars that have never had an encounter within 2000 au, and the discs of stellar systems in general (i.e. single and multiple systems). The discs of protostellar systems in the 3 times solar metallicity calculation have a median characteristic radius of $\approx 65$ au and median of $\approx 20$ au in the 1/100 time solar metallicity calculation. 
    \item Discs of protostars that are isolated (they do not have a companion protostar within within 2000 au at the time they are observed) and discs in protostellar systems tend to be slightly more massive with higher metallicity. However, this relation is reversed for discs of protostars that have never had a companion within 2000 au -- discs at the lowest metallicity ($Z=0.01~\text{Z}_\odot$) are typically twice as massive as the highest metallicity ($Z=3~\text{Z}_\odot$) discs.
    \item Discs in bound pairs (either binaries or components of high-order systems) tends to be well aligned for a semi-major axes $a \lesssim 100$ au across all calculations. In the 1/100 solar metallicity calculation the discs in wide pairs $a \gtrsim 100$ are less well aligned than in the calculations with a higher metallicity with more or less random orientations.
    \item In pairs with separations $a \lesssim 100$ au discs tend to be well aligned with the orbital plane of the pairs. Again, discs and orbits in pairs with $a \gtrsim 100$ in the 1/100 solar metallicity calculation, are generally poorly aligned with essentially a random distribution of orientations between zero and 130 degrees.
    \item Relative orientation between bound sink particle spins (representative of the angular momentum of the protostar and inner part of the disc) are also sensitive to a change in metallicity. For the super-solar and solar metallicity calculations it is only those pairs closely separated ($a \lesssim 30$au) that tend to be well aligned. For the two sub-solar metallicity calculations spins of pairs separated by $a \lesssim 100$ au tend to be well aligned, but the spins of wider systems are less well aligned than for higher metallicities with essentially random alignment.  For the particular case of pairs with  separations $a > 1000$ au in the 1/100 solar metallicity calculation most of the pairs have retrograde spins relative to each other.
    \item Discs and sink particle spins of individual protostars in bound pairs tend to be  well aligned in all calculations with no significant dependence on separation. The lowest metallicity calculation has the strongest preference for alignment with $\approx 90 \%$ of discs and spins having a relative orientation angle of $45\degr$ or less. 
    \item There is no overall preferential orientation for discs in each simulation as a whole.  This is expected as magnetic fields are not included and the overall net rotation of each molecular cloud is very small.
    \item The reason that discs sizes tend to be smaller, the fractions of unresolved or low-mass discs are higher, and that the discs, orbits and protostellar spins of wide pairs tend to be less well aligned as the metallicity is decreased is because lower metallicity increases the rate of cooling of high-density gas.  This produces more gravitational fragmentation (both of collapsing molecular cloud cores and massive discs), leading to higher multiplicity and a greater role of dynamical interactions between protostars \citep{bate_metallicity_2019MNRAS.484.2341B}.
    \item The masses of the discs of the systems in the solar metallicity calculation have a similar distribution to the masses of Class 0/I discs observed in the Perseus and Orion star-forming regions. The disc mass distribution from the simulation lies in between the Class 0 disc mass distributions obtained from Perseus and Orion, is similar to the Class I distribution in Perseus, and is roughly 0.6 dex more massive than the Class I discs in the Orion star-forming.
    \item The distribution of disc radii of systems in the solar metallicity calculation is in very good agreement with the disc radius distributions of Class 0/I discs observed in the Orion star-forming region. 
    \item When compared to observations of Class II objects, the disc masses from the solar metallicity calculation are typically one order of magnitude greater than those estimated from dust observations of the discs in most local star-forming regions, $\lambda$ Ori excepted. The  radii of the discs are similarly distributed to discs observed in the Lupus region.
\end{enumerate}

In the future, the radiation hydrodynamical calculations could be improved by increasing the numerical resolution to resolve lower mass and smaller discs.  Furthermore, although the calculations include radiative transfer, separate gas and dust temperatures, and a thermochemical model of the diffuse interstellar medium, they do not include radiative feedback from the protostars themselves.  Also, magnetic fields are not included.  Despite these limitations, the disc mass and size distributions that are obtained from the calculations are in spectacularly good agreement with those of observed Class 0/I protostars in Perseus and Orion.  In particular, this seems to indicate that magnetic braking does not play a large role in determining the properties of protostellar discs \citep[c.f.][]{Seifried_etal2013,WurBatPri2019}.

\section*{Acknowledgements}
DE thanks \L{}.Tychoniec for providing a useful discussion that led to improvements to the paper.  DE acknowledges the support of a Science \& Technology Facilities Council (STFC) studentship.  This work was supported by the European Research Council under the European Commission’s Seventh Framework Programme (FP7/2007-2013 Grant Agreement No. 339248). The calculations discussed in this paper were performed on the University of Exeter Supercomputer, Isca, and on the DiRAC Complexity system, operated by the University of Leicester IT Services, which forms part of the STFC DiRAC HPC Facility (www.dirac.ac.uk). The latter equipment is funded by BIS National E-Infrastructure capital grant ST/K000373/1 and STFC DiRAC Operations grant ST/K0003259/1. DiRAC is part of the National e-Infrastructure.

\section*{Data Availability}

 {\bf SPH output files.} The data set consisting of the output from the radiation hydrodynamical calculations of \cite{bate_metallicity_2019MNRAS.484.2341B} that are analysed in this paper is available from the University of Exeter's Open Research Exeter (ORE) repository and can be accessed via the handle: http://hdl.handle.net/10871/35993.

\noindent
{\bf Figure data.} The data used to create Figs. \ref{fig:cdists_disc} to \ref{fig:disc_spin} is provided with the paper ``Supplementary Data''.



\bibliographystyle{mnras}
\bibliography{refs} 

\begin{thebibliography}{}
\makeatletter
\relax
\def\mn@urlcharsother{\let\do\@makeother \do\$\do\&\do\#\do\^\do\_\do\%\do\~}
\def\mn@doi{\begingroup\mn@urlcharsother \@ifnextchar [ {\mn@doi@}
  {\mn@doi@[]}}
\def\mn@doi@[#1]#2{\def\@tempa{#1}\ifx\@tempa\@empty \href
  {http://dx.doi.org/#2} {doi:#2}\else \href {http://dx.doi.org/#2} {#1}\fi
  \endgroup}
\def\mn@eprint#1#2{\mn@eprint@#1:#2::\@nil}
\def\mn@eprint@arXiv#1{\href {http://arxiv.org/abs/#1} {{\tt arXiv:#1}}}
\def\mn@eprint@dblp#1{\href {http://dblp.uni-trier.de/rec/bibtex/#1.xml}
  {dblp:#1}}
\def\mn@eprint@#1:#2:#3:#4\@nil{\def\@tempa {#1}\def\@tempb {#2}\def\@tempc
  {#3}\ifx \@tempc \@empty \let \@tempc \@tempb \let \@tempb \@tempa \fi \ifx
  \@tempb \@empty \def\@tempb {arXiv}\fi \@ifundefined
  {mn@eprint@\@tempb}{\@tempb:\@tempc}{\expandafter \expandafter \csname
  mn@eprint@\@tempb\endcsname \expandafter{\@tempc}}}

\bibitem[\protect\citeauthoryear{{Aizawa}, {Suto}, {Oya}, {Ikeda}  \&
  {Nakazato}}{{Aizawa} et~al.}{2020}]{masataka_2020_alignment}
{Aizawa} M.,  {Suto} Y.,  {Oya} Y.,  {Ikeda} S.,   {Nakazato} T.,  2020,
  \mn@doi [\apj] {10.3847/1538-4357/aba43d}, \href
  {https://ui.adsabs.harvard.edu/abs/2020ApJ...899...55A} {899, 55}

\bibitem[\protect\citeauthoryear{{Andre}, {Ward-Thompson}  \&
  {Barsony}}{{Andre} et~al.}{1993}]{andre_class_0_1993}
{Andre} P.,  {Ward-Thompson} D.,   {Barsony} M.,  1993, \mn@doi [\apj]
  {10.1086/172425}, \href
  {https://ui.adsabs.harvard.edu/abs/1993ApJ...406..122A} {406, 122}

\bibitem[\protect\citeauthoryear{{Ansdell} et~al.,}{{Ansdell}
  et~al.}{2016}]{ansdell_2016_lupus}
{Ansdell} M.,  et~al., 2016, \mn@doi [\apj] {10.3847/0004-637X/828/1/46}, \href
  {https://ui.adsabs.harvard.edu/abs/2016ApJ...828...46A} {828, 46}

\bibitem[\protect\citeauthoryear{{Ansdell} et~al.,}{{Ansdell}
  et~al.}{2020}]{ansdell_2020_lambda_ori}
{Ansdell} M.,  et~al., 2020, \mn@doi [\aj] {10.3847/1538-3881/abb9af}, \href
  {https://ui.adsabs.harvard.edu/abs/2020AJ....160..248A} {160, 248}

\bibitem[\protect\citeauthoryear{{Aso} et~al.,}{{Aso}
  et~al.}{2017}]{aso_2017_l1527irs}
{Aso} Y.,  et~al., 2017, \mn@doi [\apj] {10.3847/1538-4357/aa8264}, \href
  {https://ui.adsabs.harvard.edu/abs/2017ApJ...849...56A} {849, 56}

\bibitem[\protect\citeauthoryear{{Barenfeld}, {Carpenter}, {Sargent}, {Isella}
  \& {Ricci}}{{Barenfeld} et~al.}{2017}]{barenfeld_2017_upp_sco}
{Barenfeld} S.~A.,  {Carpenter} J.~M.,  {Sargent} A.~I.,  {Isella} A.,
  {Ricci} L.,  2017, \mn@doi [\apj] {10.3847/1538-4357/aa989d}, \href
  {https://ui.adsabs.harvard.edu/abs/2017ApJ...851...85B} {851, 85}

\bibitem[\protect\citeauthoryear{{Bate}}{{Bate}}{2009}]{Bate2009b}
{Bate} M.~R.,  2009, \mn@doi [\mnras] {10.1111/j.1365-2966.2008.14165.x}, \href
  {http://cdsads.u-strasbg.fr/abs/2009MNRAS.392.1363B} {392, 1363}

\bibitem[\protect\citeauthoryear{Bate}{Bate}{2012}]{bate_2012_10.1111/j.1365-2966.2011.19955.x}
Bate M.~R.,  2012, \mn@doi [MNRAS] {10.1111/j.1365-2966.2011.19955.x}, 419,
  3115

\bibitem[\protect\citeauthoryear{{Bate}}{{Bate}}{2014}]{bate_metallicity_2014MNRAS.442..285B}
{Bate} M.~R.,  2014, \mn@doi [\mnras] {10.1093/mnras/stu795}, \href
  {https://ui.adsabs.harvard.edu/abs/2014MNRAS.442..285B} {442, 285}

\bibitem[\protect\citeauthoryear{{Bate}}{{Bate}}{2018}]{bate_diversity_of_discs_2018MNRAS.475.5618B}
{Bate} M.~R.,  2018, \mn@doi [\mnras] {10.1093/mnras/sty169}, \href
  {https://ui.adsabs.harvard.edu/abs/2018MNRAS.475.5618B} {475, 5618}

\bibitem[\protect\citeauthoryear{{Bate}}{{Bate}}{2019}]{bate_metallicity_2019MNRAS.484.2341B}
{Bate} M.~R.,  2019, \mn@doi [\mnras] {10.1093/mnras/stz103}, \href
  {https://ui.adsabs.harvard.edu/abs/2019MNRAS.484.2341B} {484, 2341}

\bibitem[\protect\citeauthoryear{{Bate} \& {Burkert}}{{Bate} \&
  {Burkert}}{1997}]{bate_resolution_1997MNRAS.288.1060B}
{Bate} M.~R.,  {Burkert} A.,  1997, \mn@doi [\mnras]
  {10.1093/mnras/288.4.1060}, \href
  {https://ui.adsabs.harvard.edu/abs/1997MNRAS.288.1060B} {288, 1060}

\bibitem[\protect\citeauthoryear{Bate \& Keto}{Bate \&
  Keto}{2015}]{bate_keto_10.1093/mnras/stv451}
Bate M.~R.,  Keto E.~R.,  2015, \mn@doi [MNRAS] {10.1093/mnras/stv451}, 449,
  2643

\bibitem[\protect\citeauthoryear{{Bate}, {Bonnell}  \& {Price}}{{Bate}
  et~al.}{1995}]{bate_sphng_1995MNRAS.277..362B}
{Bate} M.~R.,  {Bonnell} I.~A.,   {Price} N.~M.,  1995, \mn@doi [MNRAS]
  {10.1093/mnras/277.2.362}, \href
  {https://ui.adsabs.harvard.edu/abs/1995MNRAS.277..362B} {277, 362}

\bibitem[\protect\citeauthoryear{{Bate}, {Bonnell}  \& {Bromm}}{{Bate}
  et~al.}{2003}]{bate_cluster_formation_2003MNRAS.339..577B}
{Bate} M.~R.,  {Bonnell} I.~A.,   {Bromm} V.,  2003, \mn@doi [\mnras]
  {10.1046/j.1365-8711.2003.06210.x}, \href
  {https://ui.adsabs.harvard.edu/abs/2003MNRAS.339..577B} {339, 577}

\bibitem[\protect\citeauthoryear{{Beckwith}, {Zuckerman}, {Skrutskie}  \&
  {Dyck}}{{Beckwith}
  et~al.}{1984}]{beckwith_discovery_of_discs_1984ApJ...287..793B}
{Beckwith} S.,  {Zuckerman} B.,  {Skrutskie} M.~F.,   {Dyck} H.~M.,  1984,
  \mn@doi [\apj] {10.1086/162738}, \href
  {https://ui.adsabs.harvard.edu/abs/1984ApJ...287..793B} {287, 793}

\bibitem[\protect\citeauthoryear{{Benz}}{{Benz}}{1990}]{benz_review_1990nmns.work..269B}
{Benz} W.,  1990, in {Buchler} J.~R.,  ed., Numerical Modelling of Nonlinear
  Stellar Pulsations Problems and Prospects. p.~269

\bibitem[\protect\citeauthoryear{{Benz}, {Bowers}, {Cameron}  \&
  {Press}}{{Benz} et~al.}{1990}]{benz_1990ApJ...348..647B}
{Benz} W.,  {Bowers} R.~L.,  {Cameron} A.~G.~W.,   {Press} W.~H.~.,  1990,
  \mn@doi [ApJ] {10.1086/168273}, \href
  {https://ui.adsabs.harvard.edu/abs/1990ApJ...348..647B} {348, 647}

\bibitem[\protect\citeauthoryear{{Boss}, {Fisher}, {Klein}  \& {McKee}}{{Boss}
  et~al.}{2000}]{boss_jeans_2000ApJ...528..325B}
{Boss} A.~P.,  {Fisher} R.~T.,  {Klein} R.~I.,   {McKee} C.~F.,  2000, \mn@doi
  [\apj] {10.1086/308160}, \href
  {https://ui.adsabs.harvard.edu/abs/2000ApJ...528..325B} {528, 325}

\bibitem[\protect\citeauthoryear{{Burke} \& {Hollenbach}}{{Burke} \&
  {Hollenbach}}{1983}]{burke_gas_grain_interaction_in_ism_1983ApJ...265..223B}
{Burke} J.~R.,  {Hollenbach} D.~J.,  1983, \mn@doi [\apj] {10.1086/160667},
  \href {https://ui.adsabs.harvard.edu/abs/1983ApJ...265..223B} {265, 223}

\bibitem[\protect\citeauthoryear{{Chavero}, {de la Reza}, {Ghezzi}, {Llorente
  de Andr{\'e}s}, {Pereira}, {Giuppone}  \& {Pinz{\'o}n}}{{Chavero}
  et~al.}{2019}]{chavero_trends_in_metallicity_2019}
{Chavero} C.,  {de la Reza} R.,  {Ghezzi} L.,  {Llorente de Andr{\'e}s} F.,
  {Pereira} C.~B.,  {Giuppone} C.,   {Pinz{\'o}n} G.,  2019, \mn@doi [\mnras]
  {10.1093/mnras/stz1496}, \href
  {https://ui.adsabs.harvard.edu/abs/2019MNRAS.487.3162C} {487, 3162}

\bibitem[\protect\citeauthoryear{{Chiaki}, {Nozawa}  \& {Yoshida}}{{Chiaki}
  et~al.}{2013}]{chiaki_dust_grain_growth_in_low_metal_cloud_2013ApJ...765L...3C}
{Chiaki} G.,  {Nozawa} T.,   {Yoshida} N.,  2013, \mn@doi [\apjl]
  {10.1088/2041-8205/765/1/L3}, \href
  {https://ui.adsabs.harvard.edu/abs/2013ApJ...765L...3C} {765, L3}

\bibitem[\protect\citeauthoryear{{Cieza} et~al.,}{{Cieza}
  et~al.}{2019}]{cieza_2019_rho_oph}
{Cieza} L.~A.,  et~al., 2019, \mn@doi [\mnras] {10.1093/mnras/sty2653}, \href
  {https://ui.adsabs.harvard.edu/abs/2019MNRAS.482..698C} {482, 698}

\bibitem[\protect\citeauthoryear{{Cunningham}, {Krumholz}, {McKee}  \&
  {Klein}}{{Cunningham}
  et~al.}{2018}]{cunningham_star_formation_2018MNRAS.476..771C}
{Cunningham} A.~J.,  {Krumholz} M.~R.,  {McKee} C.~F.,   {Klein} R.~I.,  2018,
  \mn@doi [\mnras] {10.1093/mnras/sty154}, \href
  {https://ui.adsabs.harvard.edu/abs/2018MNRAS.476..771C} {476, 771}

\bibitem[\protect\citeauthoryear{Davidson-Pilon et~al.,}{Davidson-Pilon
  et~al.}{2021}]{cameron_davidson_pilon_2021_4816284}
Davidson-Pilon C.,  et~al., 2021, CamDavidsonPilon/lifelines: 0.26.0,
  \mn@doi{10.5281/zenodo.4816284}, \url
  {https://doi.org/10.5281/zenodo.4816284}

\bibitem[\protect\citeauthoryear{{Dopcke}, {Glover}, {Clark}  \&
  {Klessen}}{{Dopcke}
  et~al.}{2011}]{dopcke_dust_cooling_on_low_metal_clouds_2011ApJ...729L...3D}
{Dopcke} G.,  {Glover} S. C.~O.,  {Clark} P.~C.,   {Klessen} R.~S.,  2011,
  \mn@doi [\apjl] {10.1088/2041-8205/729/1/L3}, \href
  {https://ui.adsabs.harvard.edu/abs/2011ApJ...729L...3D} {729, L3}

\bibitem[\protect\citeauthoryear{{Dopcke}, {Glover}, {Clark}  \&
  {Klessen}}{{Dopcke}
  et~al.}{2013}]{dopcke_dust_cooling_low_metal_2013ApJ...766..103D}
{Dopcke} G.,  {Glover} S. C.~O.,  {Clark} P.~C.,   {Klessen} R.~S.,  2013,
  \mn@doi [\apj] {10.1088/0004-637X/766/2/103}, \href
  {https://ui.adsabs.harvard.edu/abs/2013ApJ...766..103D} {766, 103}

\bibitem[\protect\citeauthoryear{{Draine}}{{Draine}}{1978}]{draine_photoelectric_1978ApJS...36..595D}
{Draine} B.~T.,  1978, \mn@doi [\apjs] {10.1086/190513}, \href
  {https://ui.adsabs.harvard.edu/abs/1978ApJS...36..595D} {36, 595}

\bibitem[\protect\citeauthoryear{{Eisner} et~al.,}{{Eisner}
  et~al.}{2018}]{eisner_2018_onc}
{Eisner} J.~A.,  et~al., 2018, \mn@doi [\apj] {10.3847/1538-4357/aac3e2}, \href
  {https://ui.adsabs.harvard.edu/abs/2018ApJ...860...77E} {860, 77}

\bibitem[\protect\citeauthoryear{{Ercolano} \& {Clarke}}{{Ercolano} \&
  {Clarke}}{2010}]{ercolano_metallicity_planet_formation_2010}
{Ercolano} B.,  {Clarke} C.~J.,  2010, \mn@doi [\mnras]
  {10.1111/j.1365-2966.2009.16094.x}, \href
  {https://ui.adsabs.harvard.edu/abs/2010MNRAS.402.2735E} {402, 2735}

\bibitem[\protect\citeauthoryear{{Fedele} et~al.,}{{Fedele}
  et~al.}{2017}]{fedele_rings_and_gaps_in_ppd_2017A&A...600A..72F}
{Fedele} D.,  et~al., 2017, \mn@doi [A\&A] {10.1051/0004-6361/201629860}, \href
  {https://ui.adsabs.harvard.edu/abs/2017A&A...600A..72F} {600, A72}

\bibitem[\protect\citeauthoryear{{Fehlberg}}{{Fehlberg}}{1969}]{fehlberg1969low}
{Fehlberg} E.,  1969, NASA Technical Report R-315

\bibitem[\protect\citeauthoryear{{Ferguson}, {Alexander}, {Allard}, {Barman},
  {Bodnarik}, {Hauschildt}, {Heffner-Wong}  \& {Tamanai}}{{Ferguson}
  et~al.}{2005}]{ferguson_opacity_tables_2005ApJ...623..585F}
{Ferguson} J.~W.,  {Alexander} D.~R.,  {Allard} F.,  {Barman} T.,  {Bodnarik}
  J.~G.,  {Hauschildt} P.~H.,  {Heffner-Wong} A.,   {Tamanai} A.,  2005,
  \mn@doi [\apj] {10.1086/428642}, \href
  {https://ui.adsabs.harvard.edu/abs/2005ApJ...623..585F} {623, 585}

\bibitem[\protect\citeauthoryear{{Fischer} \& {Valenti}}{{Fischer} \&
  {Valenti}}{2005}]{FisVal2005}
{Fischer} D.~A.,  {Valenti} J.,  2005, \mn@doi [\apj] {10.1086/428383}, \href
  {https://ui.adsabs.harvard.edu/abs/2005ApJ...622.1102F} {622, 1102}

\bibitem[\protect\citeauthoryear{{Glover} \& {Clark}}{{Glover} \&
  {Clark}}{2012a}]{glover_molecular_gas_necessary_for_star_formation_2012MNRAS.421....9G}
{Glover} S. C.~O.,  {Clark} P.~C.,  2012a, \mn@doi [\mnras]
  {10.1111/j.1365-2966.2011.19648.x}, \href
  {https://ui.adsabs.harvard.edu/abs/2012MNRAS.421....9G} {421, 9}

\bibitem[\protect\citeauthoryear{{Glover} \& {Clark}}{{Glover} \&
  {Clark}}{2012b}]{glover_metal_poor_gas_2012MNRAS.426..377G}
{Glover} S. C.~O.,  {Clark} P.~C.,  2012b, \mn@doi [\mnras]
  {10.1111/j.1365-2966.2012.21737.x}, \href
  {https://ui.adsabs.harvard.edu/abs/2012MNRAS.426..377G} {426, 377}

\bibitem[\protect\citeauthoryear{{Glover}, {Federrath}, {Mac Low}  \&
  {Klessen}}{{Glover}
  et~al.}{2010}]{glover_modelling_co_formation_2010MNRAS.404....2G}
{Glover} S.~C.~O.,  {Federrath} C.,  {Mac Low} M.~M.,   {Klessen} R.~S.,  2010,
  \mn@doi [\mnras] {10.1111/j.1365-2966.2009.15718.x}, \href
  {https://ui.adsabs.harvard.edu/abs/2010MNRAS.404....2G} {404, 2}

\bibitem[\protect\citeauthoryear{{Goldsmith}}{{Goldsmith}}{2001}]{goldsmith_thermal_balance_in_dark_cloudS_2001ApJ...557..736G}
{Goldsmith} P.~F.,  2001, \mn@doi [\apj] {10.1086/322255}, \href
  {https://ui.adsabs.harvard.edu/abs/2001ApJ...557..736G} {557, 736}

\bibitem[\protect\citeauthoryear{{Gonzalez}}{{Gonzalez}}{1997}]{Gonzalez1997}
{Gonzalez} G.,  1997, \mn@doi [\mnras] {10.1093/mnras/285.2.403}, \href
  {https://ui.adsabs.harvard.edu/abs/1997MNRAS.285..403G} {285, 403}

\bibitem[\protect\citeauthoryear{{Guthrie}}{{Guthrie}}{1985}]{Guthrie1985}
{Guthrie} B.~N.~G.,  1985, \mn@doi [\mnras] {10.1093/mnras/215.4.545}, \href
  {https://ui.adsabs.harvard.edu/abs/1985MNRAS.215..545G} {215, 545}

\bibitem[\protect\citeauthoryear{{Hale}}{{Hale}}{1994}]{Hale1994}
{Hale} A.,  1994, \mn@doi [\aj] {10.1086/116855}, \href
  {https://ui.adsabs.harvard.edu/abs/1994AJ....107..306H} {107, 306}

\bibitem[\protect\citeauthoryear{{Hollenbach} \& {McKee}}{{Hollenbach} \&
  {McKee}}{1989}]{hollenbach_molecule_formation_1989ApJ...342..306H}
{Hollenbach} D.,  {McKee} C.~F.,  1989, \mn@doi [\apj] {10.1086/167595}, \href
  {https://ui.adsabs.harvard.edu/abs/1989ApJ...342..306H} {342, 306}

\bibitem[\protect\citeauthoryear{{Hubber}, {Goodwin}  \& {Whitworth}}{{Hubber}
  et~al.}{2006}]{hubber_resolution_2006A&A...450..881H}
{Hubber} D.~A.,  {Goodwin} S.~P.,   {Whitworth} A.~P.,  2006, \mn@doi [\aap]
  {10.1051/0004-6361:20054100}, \href
  {https://ui.adsabs.harvard.edu/abs/2006A&A...450..881H} {450, 881}

\bibitem[\protect\citeauthoryear{{Ida} \& {Lin}}{{Ida} \&
  {Lin}}{2004}]{IdaLin2004}
{Ida} S.,  {Lin} D.~N.~C.,  2004, \mn@doi [\apj] {10.1086/424830}, \href
  {https://ui.adsabs.harvard.edu/abs/2004ApJ...616..567I} {616, 567}

\bibitem[\protect\citeauthoryear{{Johnson}, {Aller}, {Howard}  \&
  {Crepp}}{{Johnson} et~al.}{2010}]{johnson_giant_planets_2010}
{Johnson} J.~A.,  {Aller} K.~M.,  {Howard} A.~W.,   {Crepp} J.~R.,  2010,
  \mn@doi [\pasp] {10.1086/655775}, \href
  {https://ui.adsabs.harvard.edu/abs/2010PASP..122..905J} {122, 905}

\bibitem[\protect\citeauthoryear{{Jones} \& {Bate}}{{Jones} \&
  {Bate}}{2018}]{JonBat2018b}
{Jones} M.~O.,  {Bate} M.~R.,  2018, \mn@doi [\mnras] {10.1093/mnras/sty1969},
  \href {https://ui.adsabs.harvard.edu/abs/2018MNRAS.480.2562J} {480, 2562}

\bibitem[\protect\citeauthoryear{{Keto} \& {Caselli}}{{Keto} \&
  {Caselli}}{2008}]{keto_chemistry_2008ApJ...683..238K}
{Keto} E.,  {Caselli} P.,  2008, \mn@doi [\apj] {10.1086/589147}, \href
  {https://ui.adsabs.harvard.edu/abs/2008ApJ...683..238K} {683, 238}

\bibitem[\protect\citeauthoryear{{Krumholz}, {Klein}  \& {McKee}}{{Krumholz}
  et~al.}{2012}]{krumholz_star_formation_2012ApJ...754...71K}
{Krumholz} M.~R.,  {Klein} R.~I.,   {McKee} C.~F.,  2012, \mn@doi [\apj]
  {10.1088/0004-637X/754/1/71}, \href
  {https://ui.adsabs.harvard.edu/abs/2012ApJ...754...71K} {754, 71}

\bibitem[\protect\citeauthoryear{{Kurosawa}, {Harries}, {Bate}  \&
  {Symington}}{{Kurosawa} et~al.}{2004}]{Kurosawaetal2004}
{Kurosawa} R.,  {Harries} T.~J.,  {Bate} M.~R.,   {Symington} N.~H.,  2004,
  \mn@doi [\mnras] {10.1111/j.1365-2966.2004.07869.x}, \href
  {http://adsabs.harvard.edu/abs/2004MNRAS.351.1134K} {351, 1134}

\bibitem[\protect\citeauthoryear{{Larson}}{{Larson}}{1969}]{larson_1969MNRAS.145..271L}
{Larson} R.~B.,  1969, \mn@doi [\mnras] {10.1093/mnras/145.3.271}, \href
  {https://ui.adsabs.harvard.edu/abs/1969MNRAS.145..271L} {145, 271}

\bibitem[\protect\citeauthoryear{{Lee}, {Hirano}, {Palau}, {Ho}, {Bourke},
  {Zhang}  \& {Shang}}{{Lee} et~al.}{2009}]{lee_2009_class_0_discs}
{Lee} C.-F.,  {Hirano} N.,  {Palau} A.,  {Ho} P. T.~P.,  {Bourke} T.~L.,
  {Zhang} Q.,   {Shang} H.,  2009, \mn@doi [\apj]
  {10.1088/0004-637X/699/2/1584}, \href
  {https://ui.adsabs.harvard.edu/abs/2009ApJ...699.1584L} {699, 1584}

\bibitem[\protect\citeauthoryear{{Machida}}{{Machida}}{2008}]{Machida2008}
{Machida} M.~N.,  2008, \mn@doi [\apjl] {10.1086/590109}, \href
  {http://adsabs.harvard.edu/abs/2008ApJ...682L...1M} {682, L1}

\bibitem[\protect\citeauthoryear{{Machida}, {Omukai}, {Matsumoto}  \&
  {Inutsuka}}{{Machida} et~al.}{2009}]{Machidaetal2009}
{Machida} M.~N.,  {Omukai} K.,  {Matsumoto} T.,   {Inutsuka} S.-I.,  2009,
  \mn@doi [\mnras] {10.1111/j.1365-2966.2009.15394.x}, \href
  {http://adsabs.harvard.edu/abs/2009MNRAS.399.1255M} {399, 1255}

\bibitem[\protect\citeauthoryear{{Mathew} \& {Federrath}}{{Mathew} \&
  {Federrath}}{2020}]{MatFed2020}
{Mathew} S.~S.,  {Federrath} C.,  2020, \mn@doi [\mnras]
  {10.1093/mnras/staa1931}, \href
  {https://ui.adsabs.harvard.edu/abs/2020MNRAS.496.5201M} {496, 5201}

\bibitem[\protect\citeauthoryear{{Mathew} \& {Federrath}}{{Mathew} \&
  {Federrath}}{2021}]{MatFed2021}
{Mathew} S.~S.,  {Federrath} C.,  2021, \mn@doi [\mnras]
  {10.1093/mnras/stab2338}, \href
  {https://ui.adsabs.harvard.edu/abs/2021MNRAS.507.2448M} {507, 2448}

\bibitem[\protect\citeauthoryear{{Matsumura}, {Brasser}  \& {Ida}}{{Matsumura}
  et~al.}{2021}]{matsumura_2021_n_body_sims_ii}
{Matsumura} S.,  {Brasser} R.,   {Ida} S.,  2021, arXiv e-prints, \href
  {https://ui.adsabs.harvard.edu/abs/2021arXiv210407271M} {p. arXiv:2104.07271}

\bibitem[\protect\citeauthoryear{{McCaughrean} \& {O'dell}}{{McCaughrean} \&
  {O'dell}}{1996}]{McCODe1996}
{McCaughrean} M.~J.,  {O'dell} C.~R.,  1996, \mn@doi [\aj] {10.1086/117934},
  \href {http://adsabs.harvard.edu/abs/1996AJ....111.1977M} {111, 1977}

\bibitem[\protect\citeauthoryear{{McCaughrean}, {Stapelfeldt}  \&
  {Close}}{{McCaughrean} et~al.}{2000}]{McCStaClo2000}
{McCaughrean} M.~J.,  {Stapelfeldt} K.~R.,   {Close} L.~M.,  2000, Protostars
  and Planets IV, \href {http://adsabs.harvard.edu/abs/2000prpl.conf..485M}
  {p.~485}

\bibitem[\protect\citeauthoryear{Moe, Kratter  \& Badenes}{Moe
  et~al.}{2018}]{Moe2018_close_binary}
Moe M.,  Kratter K.~M.,   Badenes C.,  2018, \mn@doi [arXiv]
  {10.3847/1538-4357/ab0d88}

\bibitem[\protect\citeauthoryear{{Morris} \& {Monaghan}}{{Morris} \&
  {Monaghan}}{1997}]{morris_1997JCoPh.136...41M}
{Morris} J.~P.,  {Monaghan} J.~J.,  1997, \mn@doi [J. Comput. Phys.]
  {10.1006/jcph.1997.5690}, \href
  {https://ui.adsabs.harvard.edu/abs/1997JCoPh.136...41M} {136, 41}

\bibitem[\protect\citeauthoryear{{Myers}, {McKee}, {Cunningham}, {Klein}  \&
  {Krumholz}}{{Myers} et~al.}{2013}]{myers_star_formation_2013ApJ...766...97M}
{Myers} A.~T.,  {McKee} C.~F.,  {Cunningham} A.~J.,  {Klein} R.~I.,
  {Krumholz} M.~R.,  2013, \mn@doi [\apj] {10.1088/0004-637X/766/2/97}, \href
  {https://ui.adsabs.harvard.edu/abs/2013ApJ...766...97M} {766, 97}

\bibitem[\protect\citeauthoryear{{Nozawa}, {Kozasa}  \& {Nomoto}}{{Nozawa}
  et~al.}{2012}]{nozawa_low_metal_clouds_2012ApJ...756L..35N}
{Nozawa} T.,  {Kozasa} T.,   {Nomoto} K.,  2012, \mn@doi [\apjl]
  {10.1088/2041-8205/756/2/L35}, \href
  {https://ui.adsabs.harvard.edu/abs/2012ApJ...756L..35N} {756, L35}

\bibitem[\protect\citeauthoryear{{O'dell}, {Wen}  \& {Hu}}{{O'dell}
  et~al.}{1993}]{odell_hst_discs_1993ApJ...410..696O}
{O'dell} C.~R.,  {Wen} Z.,   {Hu} X.,  1993, \mn@doi [\apj] {10.1086/172786},
  \href {https://ui.adsabs.harvard.edu/abs/1993ApJ...410..696O} {410, 696}

\bibitem[\protect\citeauthoryear{{Offner}, {Klein}, {McKee}  \&
  {Krumholz}}{{Offner} et~al.}{2009}]{Offneretal2009}
{Offner} S.~S.~R.,  {Klein} R.~I.,  {McKee} C.~F.,   {Krumholz} M.~R.,  2009,
  \mn@doi [\apj] {10.1088/0004-637X/703/1/131}, \href
  {http://adsabs.harvard.edu/abs/2009ApJ...703..131O} {703, 131}

\bibitem[\protect\citeauthoryear{Offner, Kratter, Matzner, Krumholz  \&
  Klein}{Offner et~al.}{2010}]{offner2010formation}
Offner S.~S.,  Kratter K.~M.,  Matzner C.~D.,  Krumholz M.~R.,   Klein R.~I.,
  2010, \aj, 725, 1485

\bibitem[\protect\citeauthoryear{{Offner}, {Robitaille}, {Hansen}, {McKee}  \&
  {Klein}}{{Offner} et~al.}{2012}]{Offner_etal2012}
{Offner} S. S.~R.,  {Robitaille} T.~P.,  {Hansen} C.~E.,  {McKee} C.~F.,
  {Klein} R.~I.,  2012, \mn@doi [\apj] {10.1088/0004-637X/753/2/98}, \href
  {https://ui.adsabs.harvard.edu/abs/2012ApJ...753...98O} {753, 98}

\bibitem[\protect\citeauthoryear{{Ohashi} et~al.,}{{Ohashi}
  et~al.}{2014}]{ohashi_2014_l1527irs}
{Ohashi} N.,  et~al., 2014, \mn@doi [\apj] {10.1088/0004-637X/796/2/131}, \href
  {https://ui.adsabs.harvard.edu/abs/2014ApJ...796..131O} {796, 131}

\bibitem[\protect\citeauthoryear{{Omukai}}{{Omukai}}{2000}]{omukai_protostellar_collapse_metallicity_2000ApJ...534..809O}
{Omukai} K.,  2000, \mn@doi [\apj] {10.1086/308776}, \href
  {https://ui.adsabs.harvard.edu/abs/2000ApJ...534..809O} {534, 809}

\bibitem[\protect\citeauthoryear{{Ostriker}, {Stone}  \& {Gammie}}{{Ostriker}
  et~al.}{2001}]{Ostriker_turbulence_2001ApJ...546..980O}
{Ostriker} E.~C.,  {Stone} J.~M.,   {Gammie} C.~F.,  2001, \mn@doi [\apj]
  {10.1086/318290}, \href
  {https://ui.adsabs.harvard.edu/abs/2001ApJ...546..980O} {546, 980}

\bibitem[\protect\citeauthoryear{{Pollack}, {McKay}  \&
  {Christofferson}}{{Pollack}
  et~al.}{1985}]{pollack_opacity_tables_1985Icar...64..471P}
{Pollack} J.~B.,  {McKay} C.~P.,   {Christofferson} B.~M.,  1985, \mn@doi
  [\icarus] {10.1016/0019-1035(85)90069-7}, \href
  {https://ui.adsabs.harvard.edu/abs/1985Icar...64..471P} {64, 471}

\bibitem[\protect\citeauthoryear{{Price} \& {Monaghan}}{{Price} \&
  {Monaghan}}{2005}]{price_2005MNRAS.364..384P}
{Price} D.~J.,  {Monaghan} J.~J.,  2005, \mn@doi [\mnras]
  {10.1111/j.1365-2966.2005.09576.x}, \href
  {https://ui.adsabs.harvard.edu/abs/2005MNRAS.364..384P} {364, 384}

\bibitem[\protect\citeauthoryear{{Price} \& {Monaghan}}{{Price} \&
  {Monaghan}}{2007}]{price_2007MNRAS.374.1347P}
{Price} D.~J.,  {Monaghan} J.~J.,  2007, \mn@doi [\mnras]
  {10.1111/j.1365-2966.2006.11241.x}, \href
  {https://ui.adsabs.harvard.edu/abs/2007MNRAS.374.1347P} {374, 1347}

\bibitem[\protect\citeauthoryear{{R{\'e}my-Ruyer} et~al.,}{{R{\'e}my-Ruyer}
  et~al.}{2014}]{remy_dust_properties_2014A&A...563A..31R}
{R{\'e}my-Ruyer} A.,  et~al., 2014, \mn@doi [\aap]
  {10.1051/0004-6361/201322803}, \href
  {https://ui.adsabs.harvard.edu/abs/2014A&A...563A..31R} {563, A31}

\bibitem[\protect\citeauthoryear{{Sakai} et~al.,}{{Sakai}
  et~al.}{2014}]{sakai_2014_l1527}
{Sakai} N.,  et~al., 2014, \mn@doi [\apjl] {10.1088/2041-8205/791/2/L38}, \href
  {https://ui.adsabs.harvard.edu/abs/2014ApJ...791L..38S} {791, L38}

\bibitem[\protect\citeauthoryear{{Santos}, {Israelian}  \& {Mayor}}{{Santos}
  et~al.}{2001}]{Santos_etal2001}
{Santos} N.~C.,  {Israelian} G.,   {Mayor} M.,  2001, \mn@doi [\aap]
  {10.1051/0004-6361:20010648}, \href
  {https://ui.adsabs.harvard.edu/abs/2001A&A...373.1019S} {373, 1019}

\bibitem[\protect\citeauthoryear{{Segura-Cox} et~al.,}{{Segura-Cox}
  et~al.}{2016}]{segure-cox_protostar_2016ApJ...817L..14S}
{Segura-Cox} D.~M.,  et~al., 2016, \mn@doi [ApJL]
  {10.3847/2041-8205/817/2/L14}, \href
  {https://ui.adsabs.harvard.edu/abs/2016ApJ...817L..14S} {817, L14}

\bibitem[\protect\citeauthoryear{{Seifried}, {Banerjee}, {Pudritz}  \&
  {Klessen}}{{Seifried} et~al.}{2013}]{Seifried_etal2013}
{Seifried} D.,  {Banerjee} R.,  {Pudritz} R.~E.,   {Klessen} R.~S.,  2013,
  \mn@doi [\mnras] {10.1093/mnras/stt682}, \href
  {http://adsabs.harvard.edu/abs/2013MNRAS.432.3320S} {432, 3320}

\bibitem[\protect\citeauthoryear{{Tanaka} \& {Omukai}}{{Tanaka} \&
  {Omukai}}{2014}]{TanOmu2014}
{Tanaka} K. E.~I.,  {Omukai} K.,  2014, \mn@doi [\mnras]
  {10.1093/mnras/stu069}, \href
  {https://ui.adsabs.harvard.edu/abs/2014MNRAS.439.1884T} {439, 1884}

\bibitem[\protect\citeauthoryear{{Tazzari} et~al.,}{{Tazzari}
  et~al.}{2017}]{2017A&A...606A..88T}
{Tazzari} M.,  et~al., 2017, \mn@doi [\aap] {10.1051/0004-6361/201730890},
  \href {https://ui.adsabs.harvard.edu/abs/2017A&A...606A..88T} {606, A88}

\bibitem[\protect\citeauthoryear{{Tobin}, {Hartmann}, {Chiang}, {Wilner},
  {Looney}, {Loinard}, {Calvet}  \& {D'Alessio}}{{Tobin}
  et~al.}{2012}]{tobin_2012_l1527irs}
{Tobin} J.~J.,  {Hartmann} L.,  {Chiang} H.-F.,  {Wilner} D.~J.,  {Looney}
  L.~W.,  {Loinard} L.,  {Calvet} N.,   {D'Alessio} P.,  2012, \mn@doi [\nat]
  {10.1038/nature11610}, \href
  {https://ui.adsabs.harvard.edu/abs/2012Natur.492...83T} {492, 83}

\bibitem[\protect\citeauthoryear{{Tobin} et~al.,}{{Tobin}
  et~al.}{2020}]{tobin_vandam_2020ApJ...890..130T}
{Tobin} J.~J.,  et~al., 2020, \mn@doi [ApJ] {10.3847/1538-4357/ab6f64}, \href
  {https://ui.adsabs.harvard.edu/abs/2020ApJ...890..130T} {890, 130}

\bibitem[\protect\citeauthoryear{{Tripathi}, {Andrews}, {Birnstiel}  \&
  {Wilner}}{{Tripathi} et~al.}{2017}]{tripathi_2017_tau_ophi}
{Tripathi} A.,  {Andrews} S.~M.,  {Birnstiel} T.,   {Wilner} D.~J.,  2017,
  \mn@doi [\apj] {10.3847/1538-4357/aa7c62}, \href
  {https://ui.adsabs.harvard.edu/abs/2017ApJ...845...44T} {845, 44}

\bibitem[\protect\citeauthoryear{{Truelove}, {Klein}, {McKee}, {Holliman},
  {Howell}  \& {Greenough}}{{Truelove}
  et~al.}{1997}]{truelove_resolution_1997ApJ...489L.179T}
{Truelove} J.~K.,  {Klein} R.~I.,  {McKee} C.~F.,  {Holliman} John~H. I.,
  {Howell} L.~H.,   {Greenough} J.~A.,  1997, \mn@doi [\apjl] {10.1086/310975},
  \href {https://ui.adsabs.harvard.edu/abs/1997ApJ...489L.179T} {489, L179}

\bibitem[\protect\citeauthoryear{{Tsuribe} \& {Omukai}}{{Tsuribe} \&
  {Omukai}}{2006}]{tsuribe_dust_fragmentation_of_low_metal_clouds_2006ApJ...642L..61T}
{Tsuribe} T.,  {Omukai} K.,  2006, \mn@doi [\apjl] {10.1086/504290}, \href
  {https://ui.adsabs.harvard.edu/abs/2006ApJ...642L..61T} {642, L61}

\bibitem[\protect\citeauthoryear{{Tychoniec} et~al.,}{{Tychoniec}
  et~al.}{2018}]{tychoniec_2018_vandam}
{Tychoniec} {\L}.,  et~al., 2018, \mn@doi [\apjs] {10.3847/1538-4365/aaceae},
  \href {https://ui.adsabs.harvard.edu/abs/2018ApJS..238...19T} {238, 19}

\bibitem[\protect\citeauthoryear{{Tychoniec} et~al.,}{{Tychoniec}
  et~al.}{2020}]{tychoniec_2020_vandam}
{Tychoniec} {\L}.,  et~al., 2020, \mn@doi [\aap] {10.1051/0004-6361/202037851},
  \href {https://ui.adsabs.harvard.edu/abs/2020A&A...640A..19T} {640, A19}

\bibitem[\protect\citeauthoryear{{Urban}, {Martel}  \& {Evans}}{{Urban}
  et~al.}{2010}]{Urban2010}
{Urban} A.,  {Martel} H.,   {Evans} Neal~J. I.,  2010, \mn@doi [\apj]
  {10.1088/0004-637X/710/2/1343}, \href
  {https://ui.adsabs.harvard.edu/abs/2010ApJ...710.1343U} {710, 1343}

\bibitem[\protect\citeauthoryear{{Vorobyov}, {Elbakyan}, {Omukai}, {Hosokawa},
  {Matsukoba}  \& {Guedel}}{{Vorobyov}
  et~al.}{2020}]{vorobyov_low_metal_accretion_2020A&A...641A..72V}
{Vorobyov} E.~I.,  {Elbakyan} V.~G.,  {Omukai} K.,  {Hosokawa} T.,  {Matsukoba}
  R.,   {Guedel} M.,  2020, \mn@doi [\aap] {10.1051/0004-6361/202038354}, \href
  {https://ui.adsabs.harvard.edu/abs/2020A&A...641A..72V} {641, A72}

\bibitem[\protect\citeauthoryear{{Weis}}{{Weis}}{1974}]{Weis1974}
{Weis} E.~W.,  1974, \mn@doi [\apj] {10.1086/152881}, \href
  {https://ui.adsabs.harvard.edu/abs/1974ApJ...190..331W} {190, 331}

\bibitem[\protect\citeauthoryear{{Whitehouse} \& {Bate}}{{Whitehouse} \&
  {Bate}}{2006}]{whitehouse_2006MNRAS.367...32W}
{Whitehouse} S.~C.,  {Bate} M.~R.,  2006, \mn@doi [\mnras]
  {10.1111/j.1365-2966.2005.09950.x}, \href
  {https://ui.adsabs.harvard.edu/abs/2006MNRAS.367...32W} {367, 32}

\bibitem[\protect\citeauthoryear{{Whitehouse}, {Bate}  \&
  {Monaghan}}{{Whitehouse} et~al.}{2005}]{whitehouse_2005MNRAS.364.1367W}
{Whitehouse} S.~C.,  {Bate} M.~R.,   {Monaghan} J.~J.,  2005, \mn@doi [\mnras]
  {10.1111/j.1365-2966.2005.09683.x}, \href
  {https://ui.adsabs.harvard.edu/abs/2005MNRAS.364.1367W} {364, 1367}

\bibitem[\protect\citeauthoryear{{Whitworth}}{{Whitworth}}{1998}]{whitworth_jeans_instab_1998MNRAS.296..442W}
{Whitworth} A.~P.,  1998, \mn@doi [\mnras] {10.1046/j.1365-8711.1998.01479.x},
  \href {https://ui.adsabs.harvard.edu/abs/1998MNRAS.296..442W} {296, 442}

\bibitem[\protect\citeauthoryear{{Wurster}, {Bate}  \& {Price}}{{Wurster}
  et~al.}{2019}]{WurBatPri2019}
{Wurster} J.,  {Bate} M.~R.,   {Price} D.~J.,  2019, \mn@doi [\mnras]
  {10.1093/mnras/stz2215}, \href
  {https://ui.adsabs.harvard.edu/abs/2019MNRAS.489.1719W} {489, 1719}

\bibitem[\protect\citeauthoryear{{Yasui}, {Kobayashi}, {Tokunaga}, {Saito}  \&
  {Tokoku}}{{Yasui} et~al.}{2009}]{yasui_low_metal_disc_lifetime_2009}
{Yasui} C.,  {Kobayashi} N.,  {Tokunaga} A.~T.,  {Saito} M.,   {Tokoku} C.,
  2009, \mn@doi [\apj] {10.1088/0004-637X/705/1/54}, \href
  {https://ui.adsabs.harvard.edu/abs/2009ApJ...705...54Y} {705, 54}

\bibitem[\protect\citeauthoryear{{Yasui}, {Kobayashi}, {Saito}  \&
  {Izumi}}{{Yasui} et~al.}{2016}]{yasui_2016_low_metal_sh_2_208}
{Yasui} C.,  {Kobayashi} N.,  {Saito} M.,   {Izumi} N.,  2016, \mn@doi [\aj]
  {10.3847/0004-6256/151/5/115}, \href
  {https://ui.adsabs.harvard.edu/abs/2016AJ....151..115Y} {151, 115}

\bibitem[\protect\citeauthoryear{Yasui, Kobayashi, Saito, Izumi  \&
  Skidmore}{Yasui et~al.}{2021}]{Yasui_et_al2021}
Yasui C.,  Kobayashi N.,  Saito M.,  Izumi N.,   Skidmore W.,  2021, \mn@doi
  [The Astronomical Journal] {10.3847/1538-3881/abd331}, 161, 139

\bibitem[\protect\citeauthoryear{{Yen}, {Takakuwa}, {Ohashi}  \& {Ho}}{{Yen}
  et~al.}{2013}]{yen_2013_class_0_discs}
{Yen} H.-W.,  {Takakuwa} S.,  {Ohashi} N.,   {Ho} P. T.~P.,  2013, \mn@doi
  [\apj] {10.1088/0004-637X/772/1/22}, \href
  {https://ui.adsabs.harvard.edu/abs/2013ApJ...772...22Y} {772, 22}

\bibitem[\protect\citeauthoryear{{Yen}, {Koch}, {Takakuwa}, {Ho}, {Ohashi}  \&
  {Tang}}{{Yen} et~al.}{2015}]{yen_class_0_0I_2015}
{Yen} H.-W.,  {Koch} P.~M.,  {Takakuwa} S.,  {Ho} P. T.~P.,  {Ohashi} N.,
  {Tang} Y.-W.,  2015, \mn@doi [\apj] {10.1088/0004-637X/799/2/193}, \href
  {https://ui.adsabs.harvard.edu/abs/2015ApJ...799..193Y} {799, 193}

\bibitem[\protect\citeauthoryear{{Yen}, {Koch}, {Takakuwa}, {Krasnopolsky},
  {Ohashi}  \& {Aso}}{{Yen}
  et~al.}{2017}]{yen_protostar_obs_2017ApJ...834..178Y}
{Yen} H.-W.,  {Koch} P.~M.,  {Takakuwa} S.,  {Krasnopolsky} R.,  {Ohashi} N.,
  {Aso} Y.,  2017, \mn@doi [ApJ] {10.3847/1538-4357/834/2/178}, \href
  {https://ui.adsabs.harvard.edu/abs/2017ApJ...834..178Y} {834, 178}

\bibitem[\protect\citeauthoryear{{Zucconi}, {Walmsley}  \& {Galli}}{{Zucconi}
  et~al.}{2001}]{zucconi_dust_temp_2001A&A...376..650Z}
{Zucconi} A.,  {Walmsley} C.~M.,   {Galli} D.,  2001, \mn@doi [\aap]
  {10.1051/0004-6361:20010778}, \href
  {https://ui.adsabs.harvard.edu/abs/2001A&A...376..650Z} {376, 650}

\bibitem[\protect\citeauthoryear{{van Terwisga}, {Hacar}  \& {van
  Dishoeck}}{{van Terwisga} et~al.}{2019}]{van_terwisga_2019_omc2}
{van Terwisga} S.~E.,  {Hacar} A.,   {van Dishoeck} E.~F.,  2019, \mn@doi
  [\aap] {10.1051/0004-6361/201935378}, \href
  {https://ui.adsabs.harvard.edu/abs/2019A&A...628A..85V} {628, A85}

\makeatother
\end{thebibliography}


\section*{Supporting Information}\label{sect:supp info}

Supplementary data are available at MNRAS online.

We provide text files containing the data required to reproduce 
Figs. \ref{fig:cdists_disc} to \ref{fig:disc_spin}.  These data files
are the equivalents of those provided by \citet{bate_diversity_of_discs_2018MNRAS.475.5618B}, but were derived
from the analysis of the radiation hydrodynamical calculations 
of \citet{bate_metallicity_2019MNRAS.484.2341B}.

{\bf Circumstellar disc data files.} One text file is provided for each sink particle in each of the calculations, that gives the time evolution of the properties of the protostar and its circumstellar disc.  The data necessary to construct Figs.~\ref{fig:cdists_disc}, \ref{fig:no enc mass}, and Fig.~\ref{fig:disc_spin} is contained in these files.  Their file names are of the format "Disc\_AAA\_UUUUUUUU.txt", where "AAA" gives the number of the sink particle in order of it formation (e.g. "001" for the first sink particle in each calculation).  The number "UUUUUUUU" gives the unique particle identification number from the {\tt sphNG} simulation.  Each line of a file contains 26 numbers delimited by spaces that give the state of the protostar at one instance in time.  The following information is given: (1) time, (2) time of formation of the protostar, (3) mass of the protostar, (4) mass of the circumstellar disc, (5-17) 13 numbers that give the radii that contain 2, 5, 10, 20, 30, 40, 50, 63.2, 70, 80, 90, 95, and 100\% of the circumstellar disc, (18) an integer which is 1 if the protostar has no companion within 2000 au and 2 if there is at least one companion, (19-21) three numbers that give the angular momentum of the circumstellar disc, (22-24) three numbers that give the spin angular momentum of the protostar, (25) an integer whose absolute value gives the number of the nearest sink particle if (18) is equal to 2 but zero otherwise, (26) an integer whose absolute value gives the unique particle identification number of the nearest sink particle.  Integers (25) and (26) are negative if the companion is not bound to the protostar.  Time is given units of $\sqrt{(0.1~{\rm pc})^3/({\rm G~M}_\odot)} = 471300$~yrs, masses are given in M$_\odot$, radii are given in au.  Angular momentum is in units of $\sqrt{({\rm G~M}_\odot^3(0.1~{\rm pc})}$.

{\bf Data files for protostellar systems.} We provide one text file for each bound system of protostars in each of the calculations.  The data necessary to construct Figs.~\ref{fig:cdists_sys} and \ref{fig:sys_mass_mass} is contained in these files.  Their file names are of the format "SysDMR\_N(\_AAA).txt", where "N" gives the number of protostars in the system and there is one occurrence of "\_AAA" for each protostar to give the numbers of the sink particles (e.g. SysDMR\_1\_001.txt or SysDMR\_4\_036\_037\_044\_049.txt).  Each line of a file contains 8 numbers delimited by spaces that give the state of the system at one instance in time.  The following information is given: (1) an integer giving the number of protostars in the system which may be 1, 2, 3, or 4, (2) time, (3) the age of the oldest protostar in the system, (4) total protostellar mass of the system, (5) the mass of the primary, (6) the total mass in all of the system's discs, (7) the characteristic disc radius that contains 50\% of the total disc mass, (8) the characteristic disc radius that contains 63.2\% of the total disc mass.  The units are the same as those used in the other data files.

{\bf Data files for bound protostellar pairs.} The format of these files differs slightly to the equivalent provided by \citet{bate_diversity_of_discs_2018MNRAS.475.5618B}.  We provide one text file for each bound pair of protostars in each of the calculations.  The data necessary to construct Figs.~\ref{fig:disc_disc} to \ref{fig:spin_spin} is contained in these files.  Their file names are of the format "Pair\_AAA\_BBB.txt", where "AAA" and "BBB" give the numbers of the two sink particles that form the pair.  Each line of a file contains 14 numbers delimited by spaces that give the state of the pair at one instance in time.  The following information is given: (1) the number of sink particles in the system containing the pair, which may be 2, 3, or 4, (2-3) the numbers of the two sink particles that form the pair, (4) the age of the oldest protostar in the pair (in years), (5) time (in code units), (6) total protostellar mass of the pair, (7) the mass of the primary, (8) the semi-major axis of the pair, (9) the relative orientation angle between the two circumstellar discs, (10-11) the relative orientation angles between the primary's disc and the orbit, and the secondary's disc and the orbit, (12) the relative orientation angle between the two protostellar spins, (13-14) the relative orientation angles between the primary's disc and its spin, and the secondary's disc and its spin.  The semi-major axis is given in au, and all angles are given in degrees.




\bsp	
\label{lastpage}
\end{document}